\documentclass[preprint,aps,showkeys,nofootinbib,11pt]{revtex4-1}
\usepackage{geometry}                % See geometry.pdf to learn the layout options. There are lots.
\usepackage{amsmath,amssymb,amsfonts,dcolumn,color,graphicx,graphics,latexsym,placeins,epsfig,tikz}
\usetikzlibrary{arrows,shapes}
\usepackage{subfigure,rotating,bm,mathrsfs}
\usepackage[title]{appendix}
\usepackage{psfrag}

\newcommand{\be}{\begin{equation}}
\newcommand{\ee}{\end{equation}}
\newcommand{\bear}{\begin{eqnarray}}
\newcommand{\eear}{\end{eqnarray}}
\newcommand{\ba}{\begin{array}}
\newcommand{\ea}{\end{array}}

\usepackage[pagebackref=false, colorlinks=true]{hyperref}
\hypersetup{linkcolor=red,         % color of internal links
                  citecolor=blue,        % color of links to bibliography
                  filecolor=magenta,      % color of file links
                  urlcolor=blue}          % color of the url
\begin{document}

\title{\Large Perfect fluid with shear viscosity and spacetime evolution}

\author{Inyong Cho}
\email{iycho@seoultech.ac.kr}
\author{Rajibul Shaikh}
\email{lrajibulsk@gmail.com}
\affiliation{Institute of Convergence Fundamental Studies, School of Natural Sciences, College of Liberal Arts, \\ Seoul National University of Science and Technology, Seoul 01811, Korea}

\begin{abstract}
We investigate the anisotropic evolution of spacetime driven by perfect fluid
with off-diagonal shear-viscosity components. We consider the simplest form of the equation of state for fluid,
for which the pressure and the shear stress are proportional to the energy density individually.
At late times, compared with the usual Friedmann universe, 
we find that the spacetime expands less rapidly as the energy density drops faster 
due to the transfer to the shear stress. 
Very interestingly, for some ranges of the equation-of-state parameters, 
we find that the initial big-bang singularity can be removed. 
\end{abstract}


\maketitle

\section{Introduction}
It is well known how the Friedmann universe evolves with perfect fluid, 
in particular, barotropic fluid.
With the energy-momentum tensor,
\be
T^\mu_\nu = {\rm diag}[-\rho,p,p,p],
\ee
of which the equation of state is given by $p=w\rho$,
and with the metric ansatz,
\be\label{metricFRW}
ds^2=-dt^2+a^2(t)\left(dx^2+dy^2+dz^2\right),
\ee
the solutions to the Einstein's equation provides
the scale factor, the energy density, and the three-volume density as (for $w>-1$)
\begin{align}
a &= a_0 t^{2/[3(1+w)]}, \label{FRWa}\\
\rho &= \rho_0 a^{-3(1+w)} \propto \frac{1}{t^2}, \label{FRWrho}\\
{\cal V}_3 &\equiv \sqrt{g^{(3)}} = a^3 = a_0^3t^{2/(1+w)}. \label{FRWV3}
\end{align}
However, the Friedmann universe with the off-diagonal stress components ($T^i_j$) of fluid
has not been studied well.
There are a few trials considering these components in fluid dynamics \citep{FLUID1,FLUID2,MTW},
but not in the scope of spacetime structure.
In this paper, we consider the fluid of which the energy-momentum tensor contains
the off-diagonal stress terms in addition to the diagonal terms.

The physical meaning of $T^{ij}$ is 
the flux of the $i$-component of momentum across the surface of $x^{j}={\rm constant}$ \cite{MTW}. Hence, it represents a stress in the $i$ direction on the surface of $x^{j}={\rm constant}$.
This type of stress component may arise in the cosmological perturbations
when both the scalar and the tensor modes are introduced.
In particular, it appears in the effective energy-momentum tensor 
composed of the quadratic terms of the coupled linear scalar and tensor modes.
The $\times$-polarization tensor mode induces 
such off-diagonal components in the energy-momentum tensor,
while the $+$-polarization mode induces diagonal components.
Although the quadratic quantities of the perturbation will be small,
the precision cosmology of gravity wave in the upcoming stage 
may perceive the effect of this type of phenomena.

In Eckart's theory, the energy-momentum tensor of fluid 
with heat flow and viscosity is given by \citep{Eckart:1940te,MTW}
\begin{equation}\label{eq:IF}
T^{\mu\nu}=\rho u^\mu u^\nu+(p-\xi\Theta)h^{\mu\nu}+q^\mu u^\nu+q^\nu u^\mu -2\eta \sigma^{\mu\nu}.
\end{equation}
Here, $\xi$ and $\eta$ are the bulk- and the shear-viscosity coefficients, 
$q^\mu$ is the heat-flux four-vector, 
$u^\mu$ is the four-velocity of fluid, 
$h_{\mu\nu}=g_{\mu\nu}+u_\mu u_\nu$ is the projection tensor, 
$\Theta=u^{\mu}_{;\mu}$ is the expansion, 
and $\sigma_{\mu\nu}=(u_{\mu;\delta}h^\delta_\nu+u_{\nu;\delta}h^\delta_\mu)/2-\Theta h_{\mu\nu}/3$ 
is the symmetric traceless shear tensor of fluid. 
The above energy-momentum tensor can be written down as the sum of three components: 
the perfect-fluid component $T^{\mu\nu}_{\rm pf}=\rho u^\mu u^\nu+p h^{\mu\nu}$, 
the heat-flux component $T^{\mu\nu}_{\rm heat}=q^\mu u^\nu+q^\nu u^\mu$, 
and the viscosity component $T^{\mu\nu}_{\rm visc}=-\xi\Theta h^{\mu\nu}-2\eta \sigma^{\mu\nu}$. 

In this work, we shall focus on the shear viscosity 
of which the energy-momentum tensor is given by
$T^{\mu\nu}_{\rm shear\; visc}=-2\eta \sigma^{\mu\nu}$. The shear tensor $\sigma^{\mu\nu}$ was derived by Eckart \citep{Eckart:1940te},
and can be expressed as above by  $u^\mu$ and $h_{\mu\nu}$.
The {\it equation of state} of  viscosity is implied in the coefficient $\eta$.
There is no simple first principle in deriving this equation of state.
Instead, it is constructed depending on the situation in a complicated way.
One example is a work by Misner \citep{Misner:1967uu}, 
in which the equation of state was derived 
by considering the effects of viscosity in the radiation,
and of anisotropic pressures from collisionless radiation.
It was obtained in a very limited situation 
for which the mean free time $\tau$ of radiation fluid is very short,
\begin{align} 
\eta = \frac{4}{15}bT^4\tau,
\end{align} 
where $b$ is the Stefan-Boltzmann constant and $T$ is the temperature.
The author was mainly interested in the anisotropy of the Universe by viscosity 
since it was right after the discovery of cosmic microwave background radiation. The same equation of state was derived in the work by Weinberg \citep{Weinberg:1971mx}.

In our work, rather than considering a specific situation as above 
to derive the equation of state, we would like to consider an {\it effective} energy-momentum tensor
to which the equation of state is imposed directly. 
As a start, we consider a simple situation for which 
the shear viscosity is proportional to the energy density, $T^i_{j\; \rm shear\; visc} = -\beta T^0_{0\; \rm pf}=\beta \rho$.
Here, $\rho$ is the energy density, and $\beta$ is an equation-of-state parameter which is an arbitrary number,
so that the result can be applied 
when one confronts a phenomenon of this type of energy-momentum tensor. 
We also restrict our concern only to the off-diagonal components of the shear tensor, 
$T^{\mu\nu}_{\rm shear\; visc}(\mu =\nu) = 0$.

Cosmology with fluid having shear viscosity has been explored in the literature in Refs.~\citep{VC1,VC2,VC3,VC4,VC5,VC6}, although there are limited exact solutions. 
In most of these works, various equations of state were imposed 
on the {\it viscosity coefficient}. The authors in Ref. \cite{VC2} studied anisotropic cosmology with $\eta\propto \rho$, although a complete exact solution has not been provided. Bianchi type-I cosmology with stiff matter ($p=\rho$) and $\eta\propto \rho^A$, where $A$ is a constant, has been studied in  Ref. \cite{VC3} and shown that, in some cases, the solutions show a new behaviour; the energy density vanishes near the initial cosmological singularity and then increases during the subsequent expansion. Banerjee et. al. \citep{VC5} studied Bianchi cosmology without assuming any equation of state on $\eta$. Instead, they considered $\sigma^{\mu\nu}\sigma_{\mu\nu}=\Theta^2$ and obtained exact solutions which lead to complicated expressions for $\eta$. Singh and Chaubey \citep{VC6} studied anisotropic cosmological models by considering $\eta\propto \rho^A$ and $\eta\propto H$, where $H$ is the average Hubble parameter. It is to be noted, therefore, that there is no first principle to consider the equation of state on $\eta$. All these different kind of equations of state were attempted in order to obtain ``exact solutions" of the field equations. 
In this work, we shall impose the equation of state directly to the energy-momentum tensor as $T^i_{j\; \rm shear\; visc} (i\neq j)\propto f(T^0_0)$, since the phenomenological situations are usually given in this way. In particular, we consider the simplest case in this work that $f(T^0_0)$ is linear function of $T^0_0$, i.e., $T^i_{j\; \rm shear\; visc} (i\neq j)\propto T^0_0$.

\clearpage
\section{Model and field equations}
\label{sec:model}
Let us consider the fluid of which the energy-momentum tensor is given by
\begin{equation}\label{eq:PF}
T^\mu_\nu=
\begin{bmatrix}
   \; -\rho \;\;\;\; &  0 \;\;\;\;\;  &  0 \;\;\;\;\;  &  0  \;\;\;\; \\
      0   &  p  & \sigma & \sigma \\
      0   & \sigma & p & \sigma \\
      0   & \sigma & \sigma &  p 
\end{bmatrix},
\end{equation}
where the energy density $\rho$, the pressure $p$, and the stress $\sigma$ depend only on time $t$. 
We introduced the {\it the off-diagonal shear viscosity} in the same way as in Ref. \cite{Misner:1967uu}. 
We assume the same shear components in order to  have the situation simplest allowing maximum symmetry.
We adopt a metric ansatz which has off-diagonal components,
\begin{align} 
ds^2 &= -dt^2+a^2(t)\left(dx^2+dy^2+dz^2\right)+2b(t)\left(dxdy+dydz+dzdx\right) \label{metricab}\\
&=-dt^2+\frac{1}{3}\left[2c^2(t)+d^2(t)\right] \left(dx^2+dy^2+dz^2\right)
+\frac{2}{3}\left[d^2(t)-c^2(t)\right]\left(dxdy+dydz+dzdx\right)\label{metriccd},
\end{align}
where we rewrote the metric \eqref{metricab} to \eqref{metriccd}
by $a^2=(2c^2+d^2)/3$ and $b=(d^2-c^2)/3$.
Later we shall see that it is easier to solve the field equations with the metric \eqref{metriccd}.
In addition, the three-volume density is given simply by ${\cal V}_3 =(a^2-b)\sqrt{a^2+2b}= c^2d$. 
For the fluid four-velocity $u^\mu=(u^t,u^x,u^y,u^z)=(1,0,0,0)$, 
the expansion and the nonzero components of the shear tensor become 
$\Theta=2\dot{c}/c+\dot{d}/d$ and $\sigma^i_j(i\neq j)=-(\dot{c}/c-\dot{d}/d)/3$. 
Then, the energy-momentum tensor \eqref{eq:PF} can be put into the form \eqref{eq:IF} 
for $\eta\neq 0$, $\xi=0$, and $q^\mu=0$, with $\sigma=2\eta(\dot{c}/c-\dot{d}/d)/3$. 

The three dimensional (3D) sector ($t$-constant hypersurface) of the metric \eqref{metricab} 
has the maximum number of Killing vectors (six for 3D space),
and the induced Riemannian metric $h_{\mu\nu}=g_{\mu\nu}+u_\mu u_\nu$
exhibits vanishing 3D Riemann curvature tensor on the hypersurface.
This indicates that the spatial sector of our metric is so called {\it maximally symmetric}, 
and the 3D space is homogeneous and isotropic \cite{Weinberg:1972kfs,Wald:1984rg}
similar to the {\it flat} FRW metric.

Although the 3D space exhibits the isotropy, 
the evolution of the spacetime is {\it anisotropic}.
By coordinate transformations,
\begin{align}\label{CT}
X = \frac{1}{\sqrt{2}} (-x+z),\quad
Y = \frac{1}{\sqrt{6}} (-x+2y-z),\quad
Z = \frac{1}{\sqrt{3}} (x+y+z),
\end{align}
the metric \eqref{metriccd} can be transformed to the Bianchi type VII metric,
\begin{align}\label{metricBianchi} 
ds^2 = -dt^2+c^2(t)(dX^2+dY^2) + d^2(t)dZ^2 ,
\end{align}
and the energy-momentum tensor \eqref{eq:PF} is transformed to
\begin{align}\label{EMTdiag}
T'^\mu_\nu =
\begin{bmatrix}
   \; -\rho \;\;\;\; &  0 \;\;\;\;\;  &  0 \;\;\;\;\;  &  0  \;\;\;\; \\
      0   &  p-\sigma  & 0 & 0 \\
      0   & 0 & p-\sigma & 0 \\
      0   & 0 & 0 &  p+2\sigma 
\end{bmatrix}.
%= {\rm diag}[-\rho,p-\sigma,p-\sigma,p+2\sigma].
\end{align}
Solving the Einstein's equation with the diagonal metric \eqref{metricBianchi} and 
the energy-momentum tensor \eqref{EMTdiag} 
is equivalent to solving with the off-diagonal ones, Eqs.~\eqref{metriccd} and  \eqref{eq:PF}.~\footnote{The physical meaning of the scale factors $c$ and $d$ 
are understood better in the diagonal metric \eqref{metricBianchi}, 
while that of the pressure $p$ and the off-diagonal shear stress $\sigma$
are understood better in the off-diagonal energy-momentum tensor \eqref{eq:PF}.}  
 
Using Eqs. \eqref{eq:PF} and \eqref{metriccd}, $tt$-, $ii$- and $ij$-component ($i\neq j$) of the Einstein's equation $G^\mu_\nu= T^\mu_\nu$ ($c=1$, $8\pi G=1$) are respectively given by
\begin{align}
\frac{\dot{c}^2}{c^2}+2\frac{\dot{c}}{c}\frac{\dot{d}}{d} &= \rho, \label{eq:FE1}\\
4\frac{\ddot{c}}{c}+2\frac{\ddot{d}}{d}+\frac{\dot{c}^2}{c^2}+2\frac{\dot{c}}{c}\frac{\dot{d}}{d} &= -3p, \label{eq:FE2}\\
\frac{\ddot{c}}{c}-\frac{\ddot{d}}{d}+\frac{\dot{c}^2}{c^2}-\frac{\dot{c}}{c}\frac{\dot{d}}{d} &= -3\sigma. \label{eq:FE3}
\end{align}
Note that, for the Friedmann universe ($c=d=a$), the shear stress $\sigma$ vanishes and the first two equations reduce to those of FRW cosmology.

In order to solve these field equations, 
we introduce two equations of state,
\be\label{eoss}
p=\alpha\rho, \qquad \sigma = \beta\rho.
\ee
With  Eqs.~\eqref{eoss} and \eqref{eq:FE1}, 
Eqs.~\eqref{eq:FE2} and \eqref{eq:FE3} can be written as
\begin{align}
2cd\ddot{c} +\frac{1+3\alpha}{2}d\dot{c}^2+(1+3\alpha)c\dot{c}\dot{d}+c^2\ddot{d} &=0, \label{eq:FE4}\\
cd\ddot{c} +(1+3\beta)d\dot{c}^2-(1-6\beta)c\dot{c}\dot{d}-c^2\ddot{d} &=0. \label{eq:FE5}
\end{align}
The addition of these equations \eqref{eq:FE4} and \eqref{eq:FE5} gives
\begin{equation}\label{eq:cDDOT}
\frac{\ddot{c}}{\dot{c}}+\frac{1+\alpha+2\beta}{2}\frac{\dot{c}}{c}+(\alpha+2\beta)\frac{\dot{d}}{d}=0,
\end{equation}
which can be integrated to give
\begin{equation}
c^{(1+\alpha+2\beta)/2}d^{\alpha+2\beta}\dot{c}=A_1 \quad \mbox{: $c$-equation},
\label{eq:cDOT}
\end{equation}
where $A_1$ is an integration constant.
Subtraction of Eq. \eqref{eq:FE5} from Eq. \eqref{eq:FE4} gives
\begin{equation}\label{d-eq}
2cd\ddot{c}+(-1+3\alpha-6\beta)d\dot{c}^2+2(2+3\alpha-6\beta)c\dot{c}\dot{d}+4c^2\ddot{d}=0,
\end{equation}
which can be rewritten as
\begin{equation}\label{eq:cdDDOT}
\frac{d}{dt}\left[c^{(-1+3\alpha-6\beta)/2}(d\dot{c}+2c\dot{d})\right]=0.
\end{equation}
Integrating this equation, we have
\begin{equation}
c^{(-1+3\alpha-6\beta)/2}(d\dot{c}+2c\dot{d})=A_2,
\label{eq:cdDOT}
\end{equation}
where $A_2$ is an integration constant.
Using Eq.~\eqref{eq:cDOT}, this finally reduces to
\begin{equation}
2c^{(1+3\alpha-6\beta)/2}\dot{d}=A_2-A_1\frac{d^{1-\alpha-2\beta}}{c^{1-\alpha+4\beta}} \quad \mbox{: $d$-equation}.
\label{eq:dDOT}
\end{equation}
Using $c$- and $d$-equation in Eqs. (\ref{eq:cDOT}) and (\ref{eq:cdDOT}), 
we obtain the energy density from Eq.~\eqref{eq:FE1},
\begin{equation}
\rho=\frac{A_1 A_2}{c^{2(1+\alpha-\beta)} d^{1+\alpha+2\beta}} \quad \mbox{: $\rho$-solution}.
\label{eq:rho}
\end{equation}
Note that $A_1$ and $A_2$ must have the same sign to make $\rho$ positive.

In this section, we manipulated the field equations and obtained the equations in simple forms;
the $c$-equation \eqref{eq:cDOT} and the $d$-equation \eqref{eq:dDOT}
which are the first-order differential equations,
and the energy density \eqref{eq:rho} in a closed form.

\section{Solutions}
\label{ssD}

In this section, we solve the field equations obtained earlier.
We present the general and special  classes depending on 
the values of the equation-of-state parameters, $\alpha$ and $\beta$.
Dividing the the $d$-equation \eqref{eq:dDOT} by the $c$-equation \eqref{eq:cDOT},
we have
\begin{equation}
2\frac{c^{\alpha-4\beta}\dot{d}}{d^{\alpha+2\beta}\dot{c}}=\frac{A_2}{A_1}-\frac{d^{1-\alpha-2\beta}}{c^{1-\alpha+4\beta}}.
\label{eq:cd_gen}
\end{equation}
Defining $x \equiv c^{1-\alpha+4\beta}$ and $y \equiv d^{1-\alpha-2\beta}$,
this equation can be written as
\begin{equation}
\frac{d}{dt}\left(x^m y\right)=\frac{A_2}{A_1}m x^m \dot{x},
\label{eq:xy_gen}
\end{equation}
where $m=(1-\alpha-2\beta)/[2(1-\alpha+4\beta)]$. 

\vspace{12pt}
\noindent
\underline{{\bf Class G}: general class}
\vspace{12pt}

Equation \eqref{eq:xy_gen} is integrated to give $d$ as a function of $c$,
\begin{equation}
d=\frac{1}{\sqrt{c}}\left[\frac{A_2}{3A_1}\frac{1-\alpha-2\beta}{1-\alpha+2\beta} c^{3(1-\alpha+2\beta)/2}
+A_3\right]^{\frac{1}{1-\alpha-2\beta}} \quad \mbox{: $d$-solution},
\label{eq:d_gen}
\end{equation}
where $A_3$ is an integration constant.
Then plugging this expression for $d$ in the $c$-equation \eqref{eq:cDOT},
we get $c$ in an implicit integral form,
\begin{equation}
t=\frac{1}{A_1}\int_{c_0}^c \sqrt{c} \left[\frac{A_2}{3A_1}\frac{1-\alpha-2\beta}{1-\alpha+2\beta} c^{3(1-\alpha+2\beta)/2}+A_3\right]^{\frac{\alpha+2\beta}{1-\alpha-2\beta}} dc   \quad \mbox{: $c$-solution},
\label{eq:t_gen}
\end{equation}
where $c_0 = c(t=0)$.
The integration can be performed numerically to obtain $c$,
and we show the result in the next section.
Now the solutions for $c$, $d$ and $\rho$ can be fully obtained 
from Eqs. \eqref{eq:t_gen}, \eqref{eq:d_gen} and \eqref{eq:rho}.

There are three classes for which Eq.~\eqref{eq:xy_gen} can not be used,
so the $d$- and $c$-solutions, \eqref{eq:d_gen} and \eqref{eq:t_gen}, are not valid;
$1-\alpha+2\beta=0$ ($m=-1$),
$1-\alpha-2\beta=0$ ($y=1$),
and $1-\alpha+4\beta=0$ ($x=1$).
These classes need to be treated separately.

\vspace{12pt}
\noindent
\underline{{\bf Class A}: $1-\alpha+2\beta=0$ ($\alpha\neq 1$ and $\beta\neq 0$)}\footnote{
If $\alpha=1$, $\beta=0$, which is the stress-free case. It will be dealt with in Class D.}
\vspace{12pt}

This class corresponds to $m=-1$.
Integrating Eq.~\eqref{eq:xy_gen}, we get
\begin{equation}
d=\frac{1}{\sqrt{c}}\left[\frac{A_2}{A_1}(1-\alpha) \log c+A_{3A}\right]^{\frac{1}{2(1-\alpha)}},
\label{eq:d_SC1}
\end{equation}
where $A_{3A}$ is an integration constant. Plugging this in the $c$-equation \eqref{eq:cDOT}, we get
\begin{equation}
t=\frac{1}{A_1}\int_{c_0}^c \sqrt{c} \left[\frac{A_2}{A_1}(1-\alpha) \log c+A_{3A}\right]^{\frac{2\alpha-1}{2(1-\alpha)}} dc.
\label{eq:t_SC1}
\end{equation}

\vspace{12pt}
\noindent
\underline{{\bf Class B}: $1-\alpha-2\beta=0$ ($\alpha\neq 1$ and $\beta\neq 0$)}
\vspace{12pt}

For this class, Eq.~\eqref{eq:cd_gen} becomes
\be
2\frac{\dot{d}}{d} = \frac{A_2}{A_1}\frac{\dot{c}}{c^{3\alpha-2}}-\frac{\dot{c}}{c},
\ee
and the solution is given by
\be
d = \frac{A_{3B}}{\sqrt{c}} \exp\left[\frac{A_2}{6(1-\alpha)A_1}c^{3(1-\alpha)}\right],
\label{eq:d_SC2}
\ee
where $A_{3B}$ is an integration constant. Plugging this in the $c$-equation \eqref{eq:cDOT}, we get
\begin{equation}
t=\frac{A_{3B}}{A_1} \int_{c_0}^c \sqrt{c} \exp\left[\frac{A_2}{6(1-\alpha)A_1}c^{3(1-\alpha)}\right] dc.
\label{eq:t_SC2}
\end{equation}

\vspace{12pt}
\noindent
\underline{{\bf Class C}: $1-\alpha+4\beta=0$ ($\alpha\neq 1$ and $\beta\neq 0$)}
\vspace{12pt}

For this class, Eq.~\eqref{eq:cd_gen} becomes
\begin{equation}
\frac{2d^{(1-3\alpha)/2}\dot{d}}{A_2/A_1-d^{3(1-\alpha)/2}} = \frac{\dot{c}}{c},
\label{eq:cd_SC3}
\end{equation}
and the solution is given by
\begin{equation}
d = \left[\frac{A_2}{A_1}-A_{3C}c^{-3(1-\alpha)/4} \right]^{\frac{2}{3(1-\alpha)}},
\label{eq:d_SC3}
\end{equation}
where $A_{3C}$ is an integration constant. Plugging this in the $c$-equation \eqref{eq:cDOT}, we get
\begin{equation}
t=\frac{1}{A_1} \int_{c_0}^c c^{(3\alpha+1)/4}
\left[\frac{A_2}{A_1}-A_{3C}c^{-3(1-\alpha)/4} \right]^{\frac{3\alpha-1}{3(1-\alpha)}} dc.
\label{eq:t_SC3}
\end{equation}

\vspace{12pt}
\noindent
\underline{{\bf Class D}: $\beta=0$}
\vspace{12pt}

This is the class for which the off-diagonal stress terms are {\it absent},
i.e., $T^i_j(i\neq j)=0$. 
The usual Friedmann universe belongs to this class.
This case is described by the general  $c$-, $d$- and $\rho$-solutions, 
Eqs. \eqref{eq:t_gen}, \eqref{eq:d_gen} and \eqref{eq:rho}.
We need to treat separately for $\alpha =1$ and $\alpha \neq1$. 
Integrating Eq. (\ref{eq:cd_gen}), we get
\begin{equation}
d=\left\{
  \begin{array}{lr}
    A_{3D} c^{(A_2 -A_1)/(2A_1)} & (\alpha =1)\\
  \frac{1}{\sqrt{c}}\left[\frac{A_2}{3A_1}c^{3(1-\alpha)/2}+A_{3D}\right]^{\frac{1}{1-\alpha}} & (\alpha \neq 1)
  \end{array}
\right.,
\label{eq:d_SC4}
\end{equation}
where $A_{3D}$ is an integration constant. Plugging this in the $c$-equation \eqref{eq:cDOT}, we get
\begin{equation}
A_1 t+A_4=\left\{
  \begin{array}{lr}
   \frac{2A_1 A_{3D}}{3A_1+A_2} c^{(3A_1+A_2)/(2A_1)} & (\alpha =1)\\
  \int \sqrt{c}\left[\frac{A_2}{3A_1}c^{3(1-\alpha)/2}+A_{3D}\right]^{\frac{\alpha}{1-\alpha}} dc & (\alpha \neq 1)
  \end{array}
\right.,
\label{eq:t_SC4}
\end{equation}
where $A_4$ is an integration constant.
%\begin{equation}
%c=\left\{
%  \begin{array}{lr}
%    \left[\frac{A_2+3A_1}{2A_1}\left(\frac{A_1}{A_3}t+A_4\right)\right]^{\frac{2A_1}{A_2+3A_1}} & : \alpha =1\\
%  \left[\frac{3(1+\alpha)}{2}\left(A_1 t+A_4\right)\right]^{\frac{2}{3(1+\alpha)}} & : \alpha \neq 1
%  \end{array}
%\right.,
%\end{equation}
%where $A_4$ is an integration constant.

{\bf (i) FRW universe}: 
Although the energy-momentum tensor is diagonal in this case, 
the metric is not since $c\neq d$ in general. 
The metric becomes diagonal and reduces to that of 
the usual Friedmann-Robertson-Walker universe when $c=d$. 
This can be achieved by setting the conditions at a certain time, say at $t=0$, 
$c=d$ and $\dot{c}=\dot{d}$. 
Applying these conditions on Eq.~\eqref{eq:d_SC4} at $t=0$, 
we find the integration constants, $A_2=3A_1$,
and $A_{3D}=1$ for $\alpha=1$ and $A_{3D}=0$ for $\alpha \neq 1$.
From Eq.~\eqref{eq:t_SC4} with these constants, we obtain
\begin{equation}
c=d=\left\{
  \begin{array}{lr}
    e^{\left(A_1 t+A_4\right)} & (\alpha =-1) \\
  \left[\frac{3(1+\alpha)}{2}\left(A_1 t+A_4\right)\right]^{\frac{2}{3(1+\alpha)}} &(\alpha \neq -1)
  \end{array}
\right.,
\label{eq:c_SC4}
\end{equation}
which makes the metric diagonal.
If we restrict the conditions further, $c(0)=c_0=1$, without loss of generality,
we get $A_4=0$ for $\alpha=-1$ and $A_4=2/[3(1+\alpha)]$ for $\alpha \neq -1$.

{\bf (ii) Non-FRW universe}:
In Sec. \ref{sec:model},
it was noted that the shear tensor is given by $\sigma^i_j(i\neq j)=-(\dot{c}/c-\dot{d}/d)/3$. 
Therefore, for the diagonal (FRW) metric in Class (i), 
the shear tensor vanished identically representing shear-free cosmology. 
However, if we do not impose the conditions $c=d$ and $\dot{c}=\dot{d}$ at $t=0$, 
the off-diagonal components of the metric are nonzero, 
and the solutions are given by Eqs. \eqref{eq:d_SC4} and \eqref{eq:t_SC4}. 
For this class, the energy-momentum tensor is diagonal, $T^i_j(i\neq j)=0$, 
while the shear tensor is nonzero,  $\sigma^i_j(i\neq j) \neq 0$. 
We shall not pay much attention to this non-FRW class in what follows.~\footnote{Note that for $\alpha=1$ with an appropriate rescaling, 
we have $c= t^{p_1}$ and $d= t^{p_3}$, 
where $p_1=2A_1/(3A_1+A_2)$ and $p_3=(A_2-A_1)/(3A_1+A_2)$. 
This is a Kasner-type solution in nonvacuum.}

\vspace{12pt}
\noindent
\underline{{\bf Kasner type}}
\vspace{12pt}

There exists an off-diagonal metric solution for the vacuum which is a Kasner-type.
From Eqs.~\eqref{eq:cdDOT} and \eqref{eq:rho},
if $A_2=0$, we have $\rho=0$ (i.e., $p=\sigma=0$) and $d=B_1/\sqrt{c}$.
From Eq.~\eqref{eq:cDOT}, we get $c=(B_2 t+B_3)^{2/3}$, where $B_i$'s are constants.
The volume density ${\cal  V}_3=B_1\left(B_2 t+B_3\right)$ becomes zero at the initial moment $t_s=-B_3/B_2$
at which the Kretshmann scalar $\mathcal{K}=64 B_2^4/[27(B_2 t+B_3)^{4}]$ diverges. 
Seen from the diagonal Bianchi-type metric \eqref{metricBianchi},
we have $c=t^{p_1}$ and $d=t^{p_3}$, where $p_1=2/3$ and $p_3=-1/3$, 
after rescaling time and setting $B_1=1$.
This is nothing but the vacuum Kasner solution \citep{Ellis1,Ellis2}.

\section{Solutions and Effect of stress}
\label{sec:solution}

In this section, we present the numerical solutions
and discuss the effect of the stress $\sigma$
by comparing with the stress-free case ($\beta=0$). 
We integrate the $c$-solution in Eq.~\eqref{eq:t_gen} 
[Eqs.~\eqref{eq:t_SC1}, \eqref{eq:t_SC2}, \eqref{eq:t_SC3}, and \eqref{eq:t_SC4} for special cases].
Then $d$ is obtained from Eq.~\eqref{eq:d_gen} 
[Eqs.~\eqref{eq:d_SC1}, \eqref{eq:d_SC2}, \eqref{eq:d_SC3}, and \eqref{eq:d_SC4} for special cases],
and $\rho$ is obtained from Eq.~\eqref{eq:rho}.

The integration constants are determined by imposing the initial conditions.
Since the original field equations are the second-order differential equations for $c$ and $d$,
we have four integration constants, $c_0$ and three $A_i$'s.
In order to compare with the the usual Friedmann universe ($\beta=0$),
we impose the initial conditions in the same way as discussed in Sec.~\ref{ssD}.

At $t=0$, we impose three conditions, $c(0)=d(0)=1$ and $\dot{c}(0)=\dot{d}(0)$. 
Then $c_0 \equiv c(0)=1$. We get $A_2/A_1=3$ from Eq.~\eqref{eq:cd_gen},
and $A_{3i}$ is obtained from Eq.~\eqref{eq:d_gen} 
[Eqs.~\eqref{eq:d_SC1}, \eqref{eq:d_SC2}, \eqref{eq:d_SC3}, and \eqref{eq:d_SC4} for special cases].
Fixing the value of $\dot{c}(0)$ in Eq.~\eqref{eq:cDOT} determines the remaining freedom $A_1$.

\begin{figure}[]
\centering
\subfigure{\includegraphics[scale=0.44]{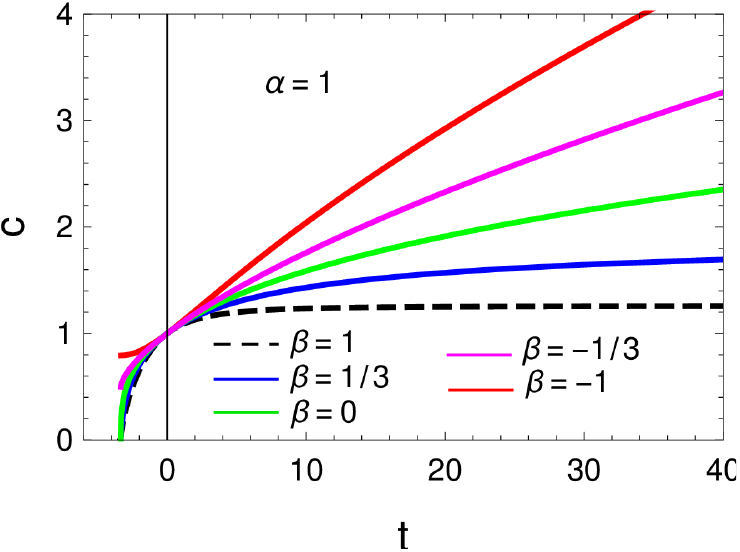}}
\subfigure{\includegraphics[scale=0.44]{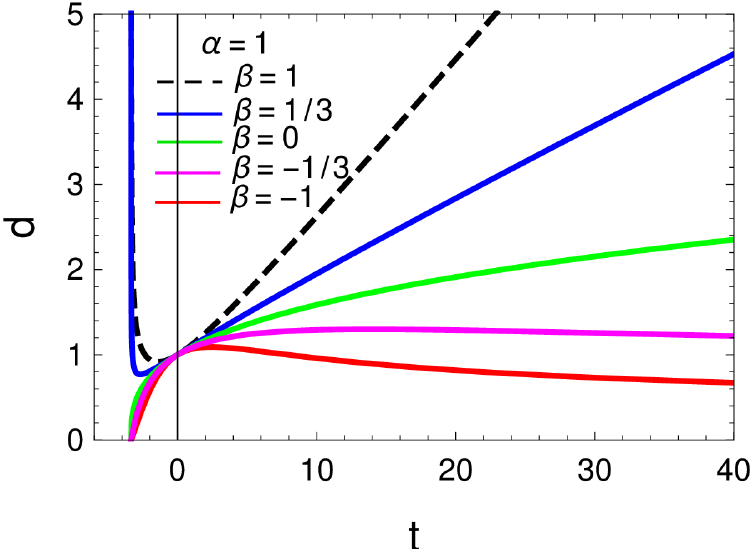}}
\subfigure{\includegraphics[scale=0.145]{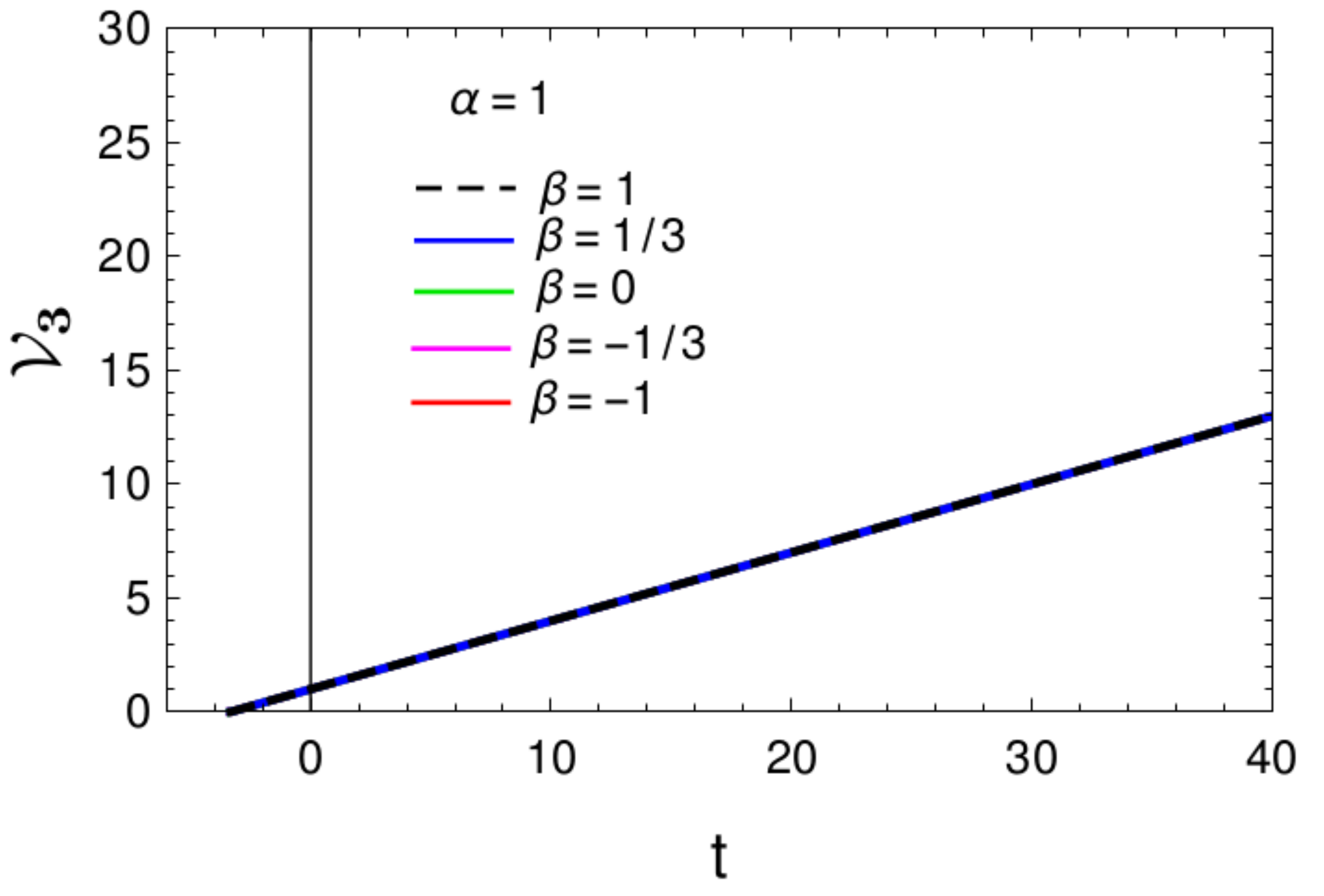}}
\subfigure{\includegraphics[scale=0.48]{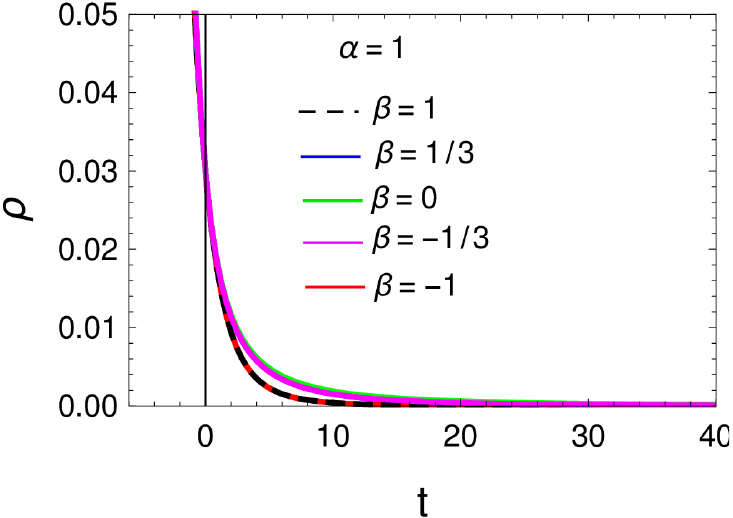}}
\subfigure{\includegraphics[scale=0.48]{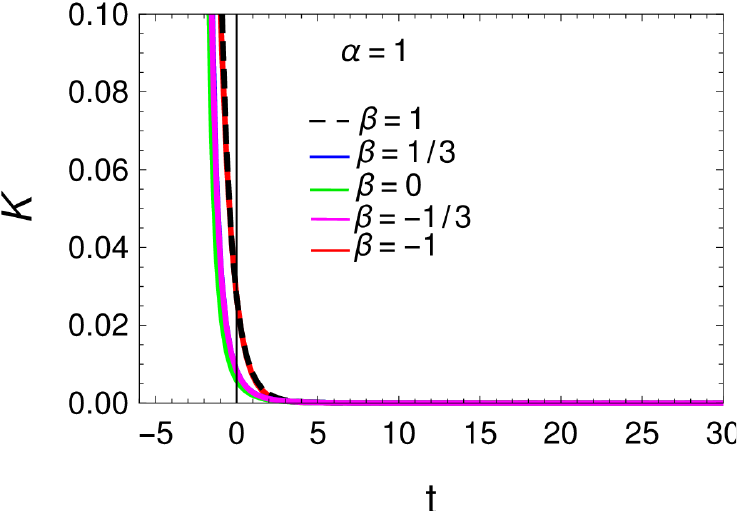}}
\subfigure{\includegraphics[scale=0.44]{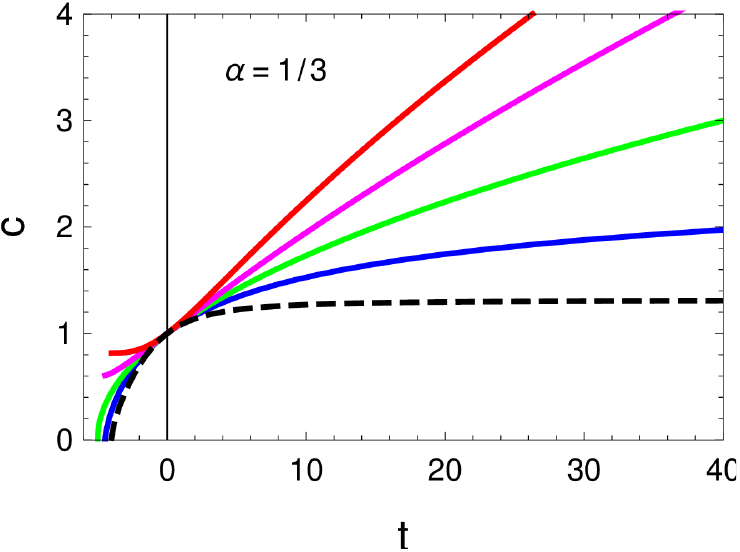}}
\subfigure{\includegraphics[scale=0.44]{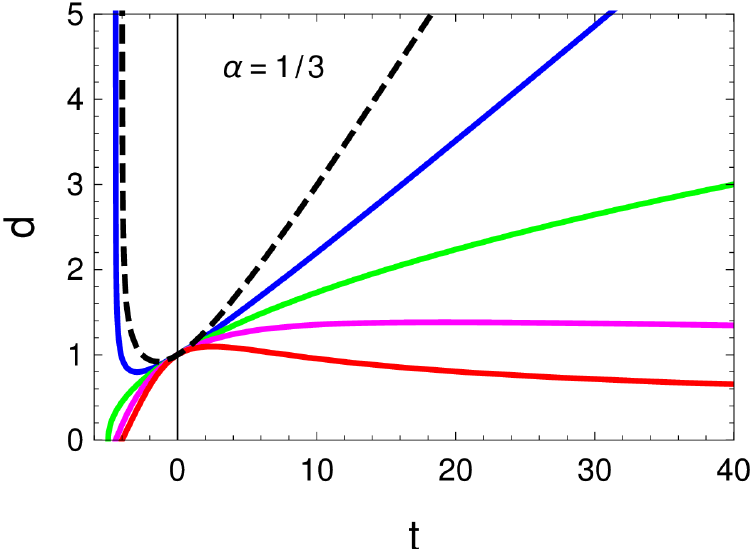}}
\subfigure{\includegraphics[scale=0.145]{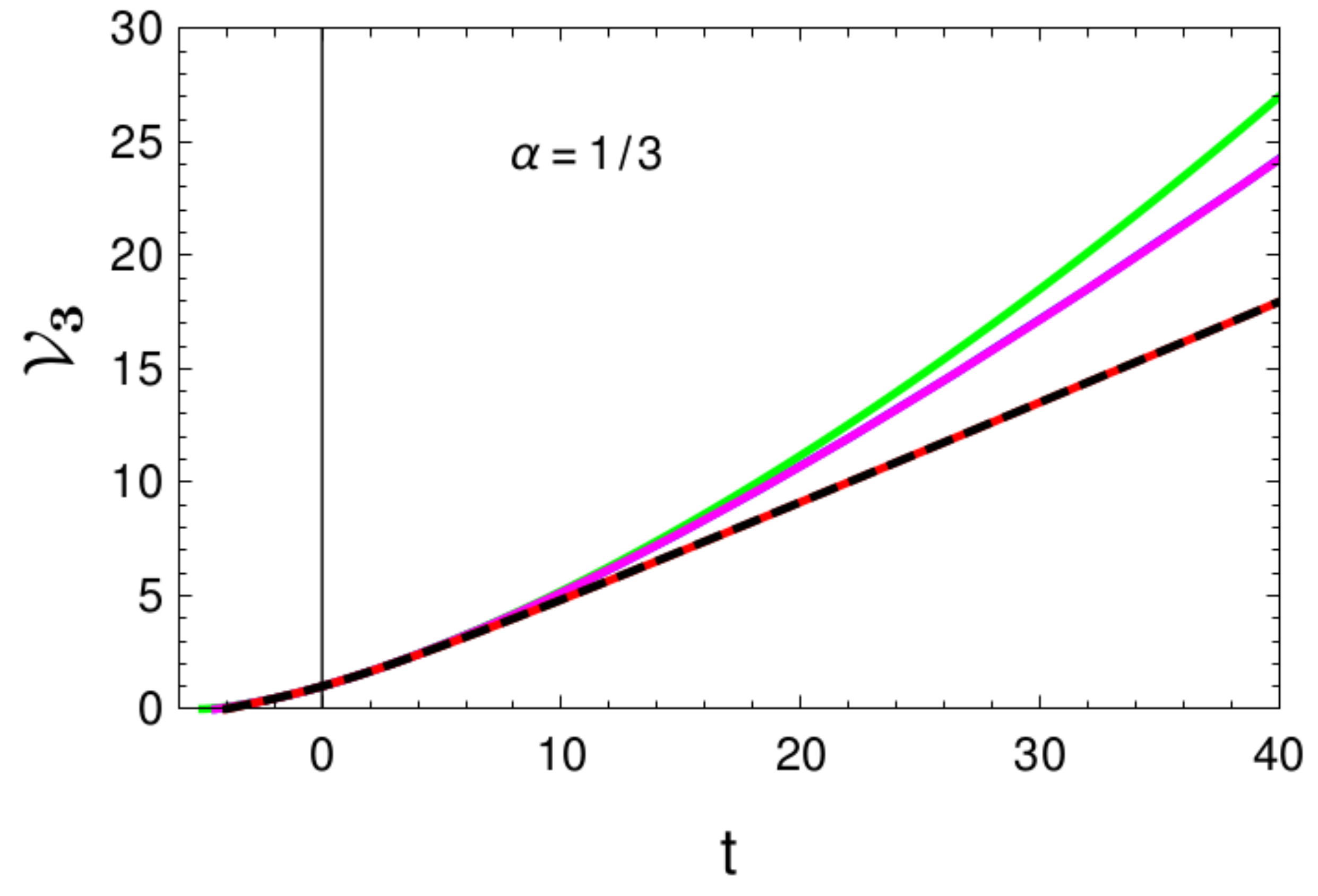}}
\subfigure{\includegraphics[scale=0.48]{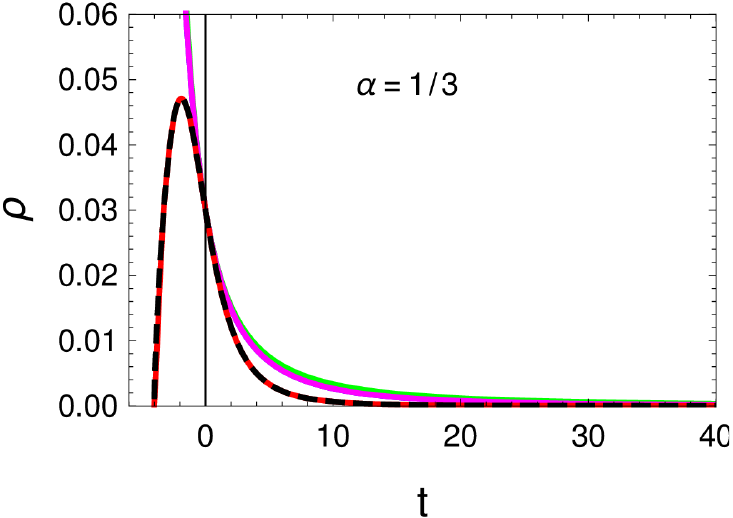}}
\subfigure{\includegraphics[scale=0.48]{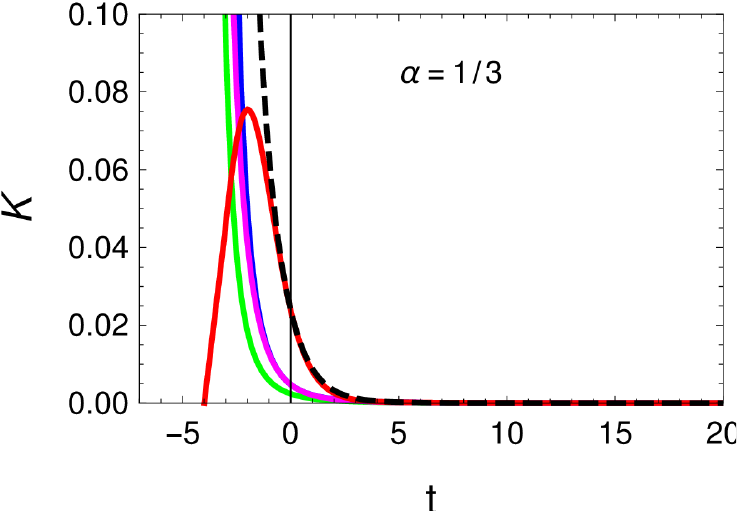}}
\subfigure{\includegraphics[scale=0.44]{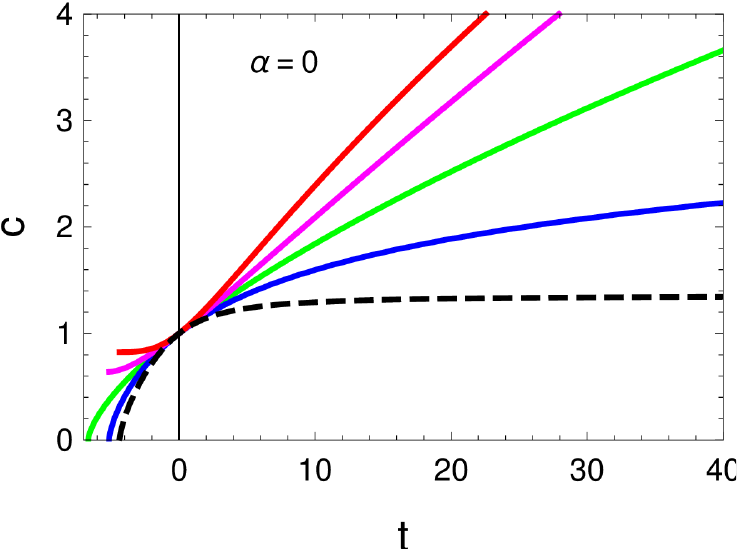}}
\subfigure{\includegraphics[scale=0.44]{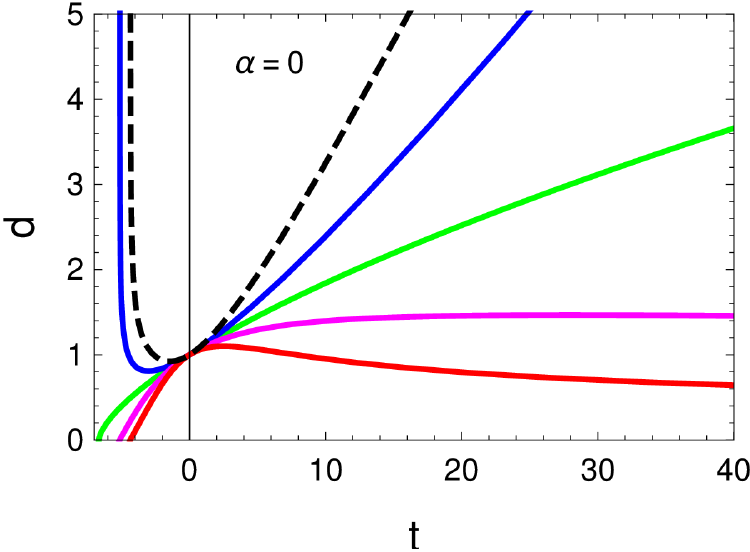}}
\subfigure{\includegraphics[scale=0.145]{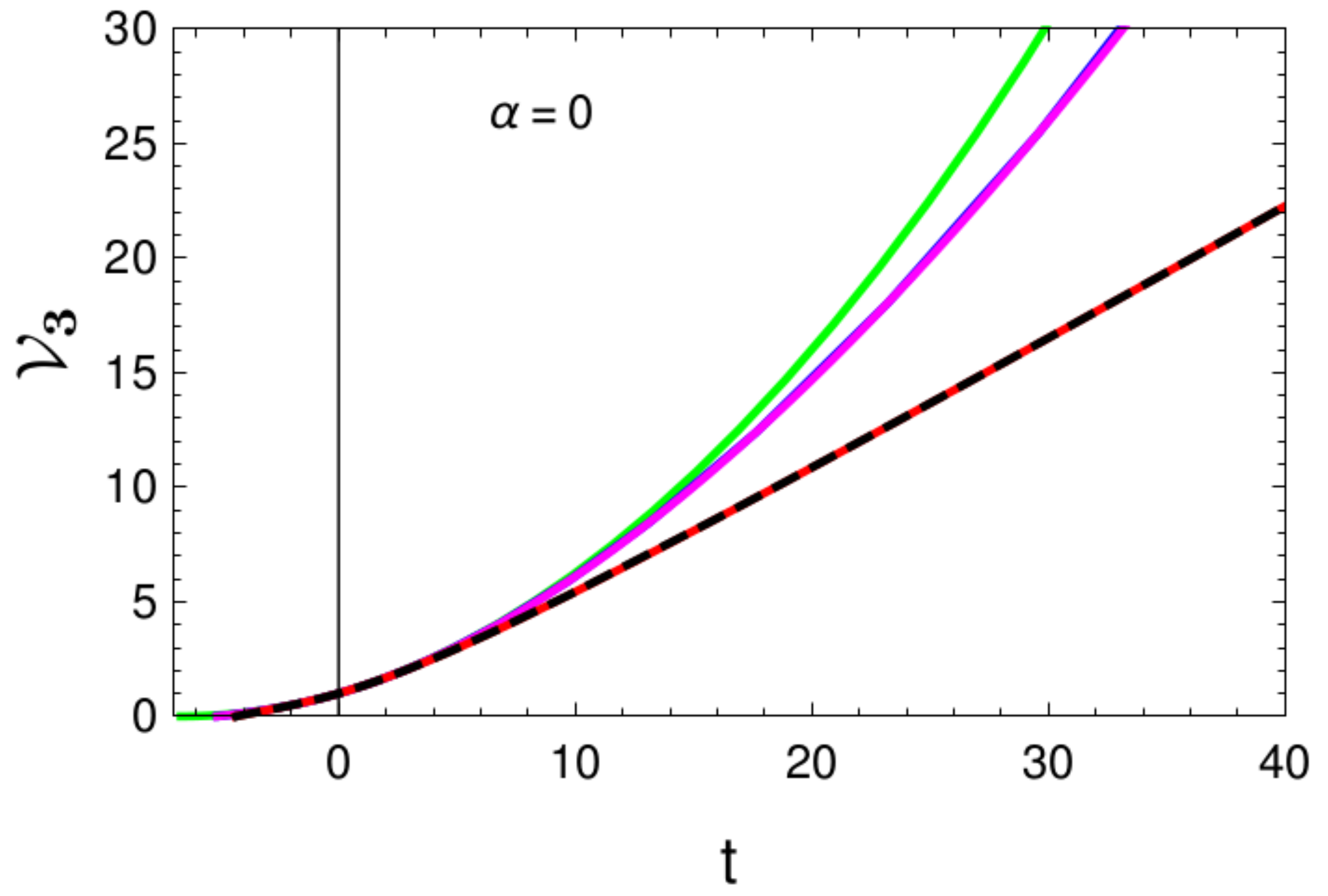}}
\subfigure{\includegraphics[scale=0.48]{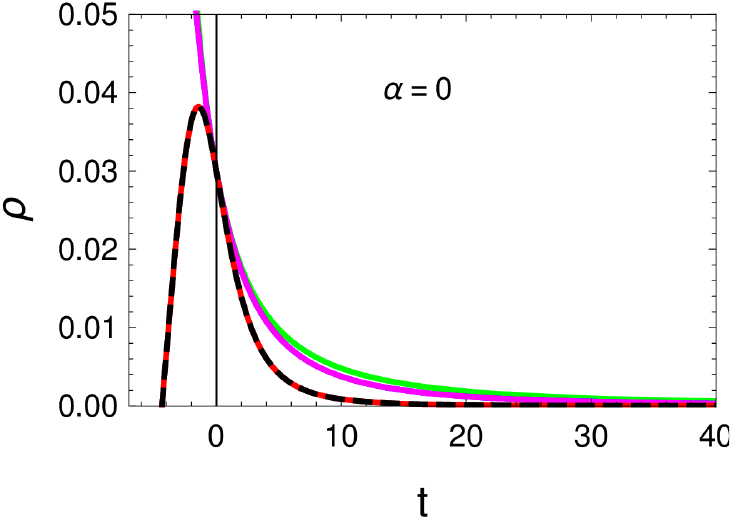}}
\subfigure{\includegraphics[scale=0.48]{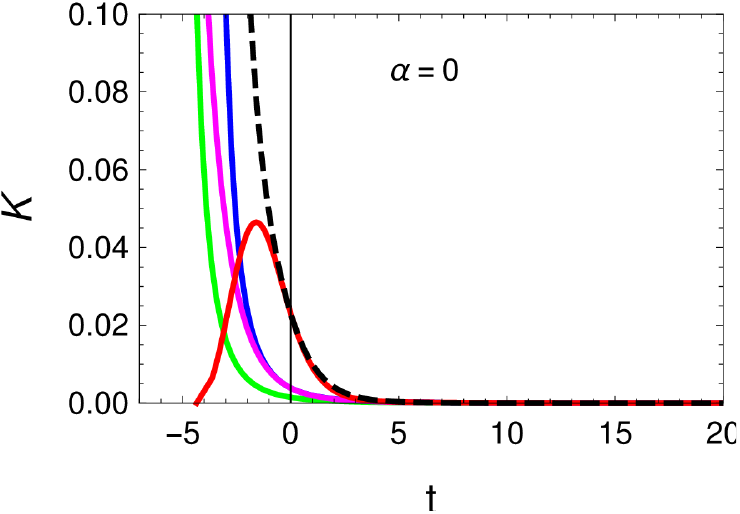}}
\subfigure{\includegraphics[scale=0.44]{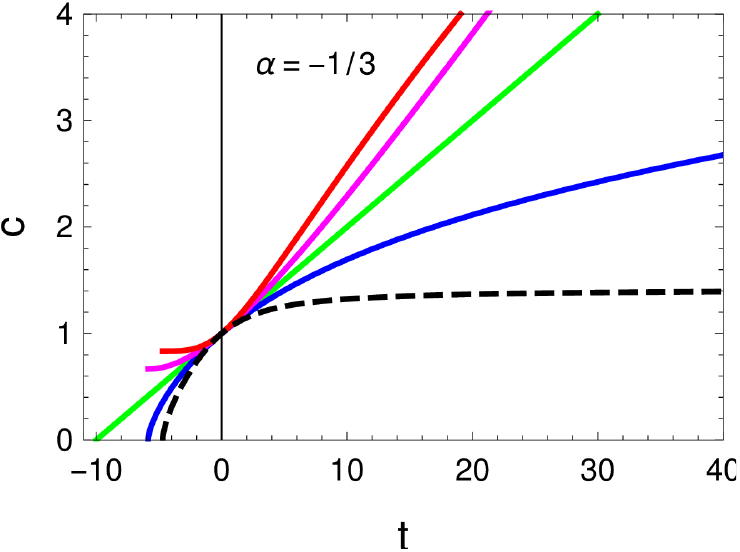}}
\subfigure{\includegraphics[scale=0.44]{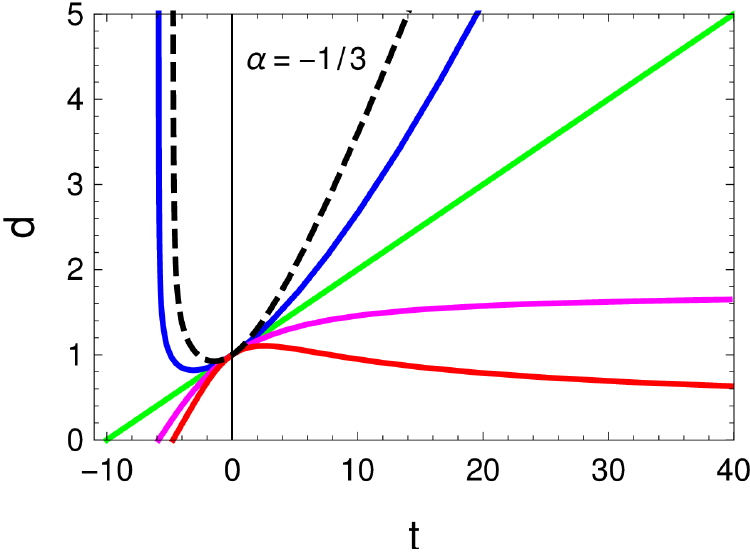}}
\subfigure{\includegraphics[scale=0.145]{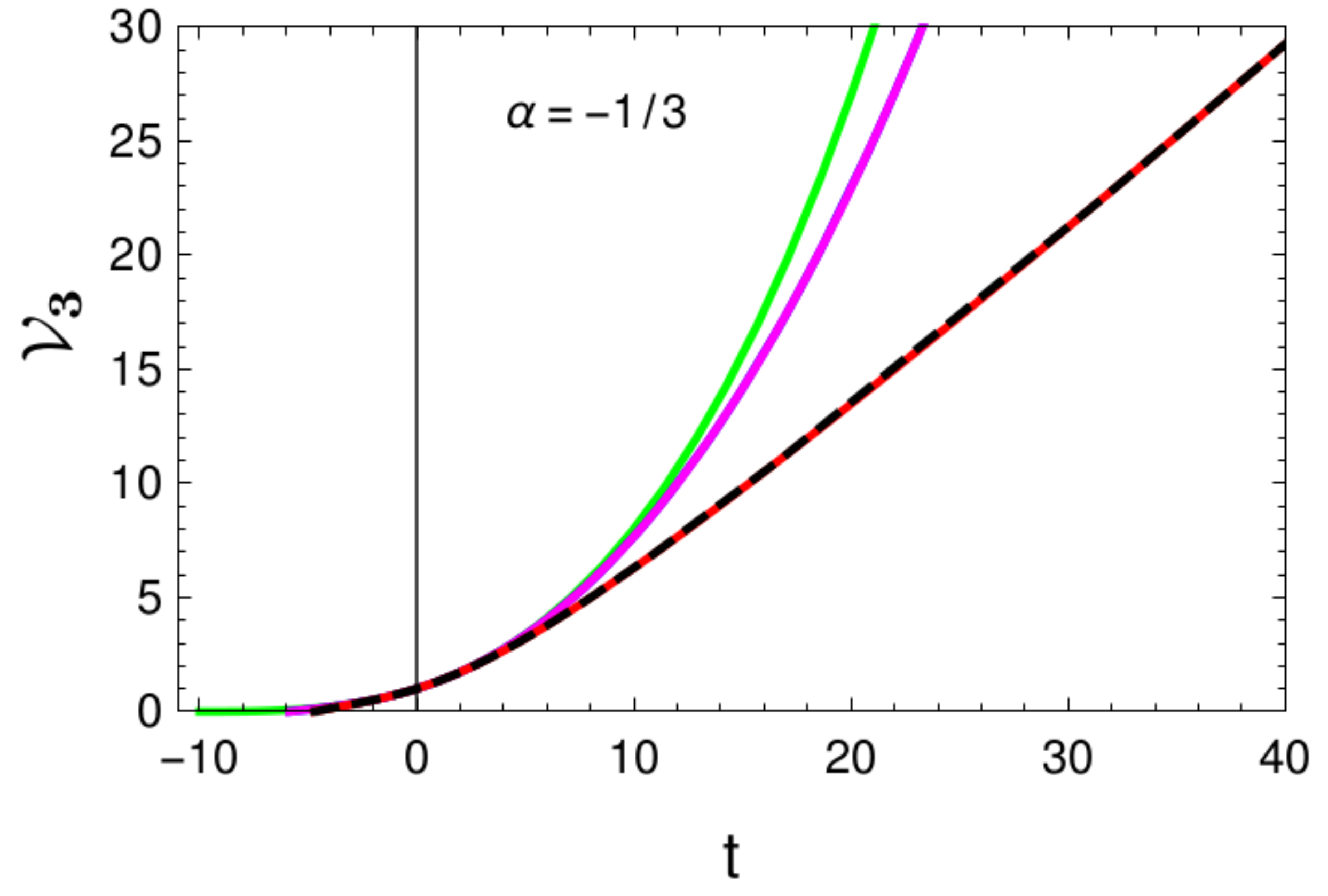}}
\subfigure{\includegraphics[scale=0.48]{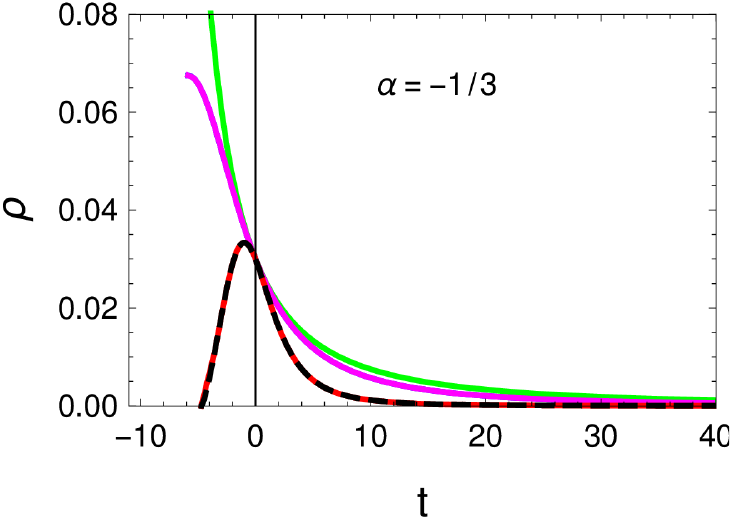}}
\subfigure{\includegraphics[scale=0.48]{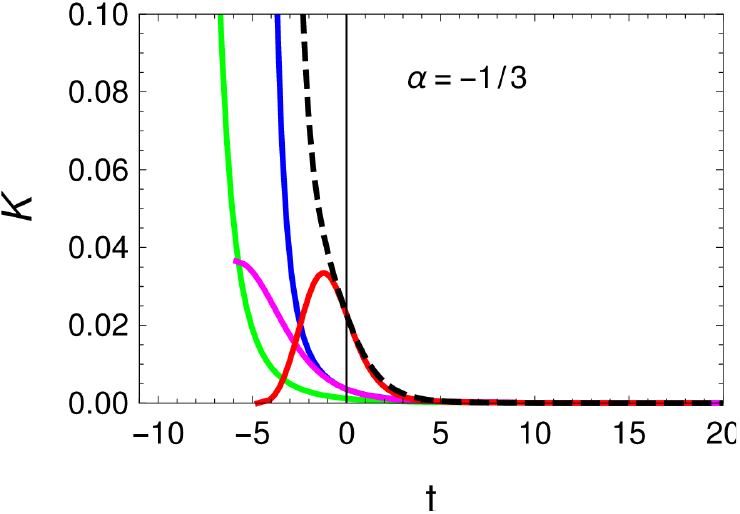}}
\subfigure{\includegraphics[scale=0.44]{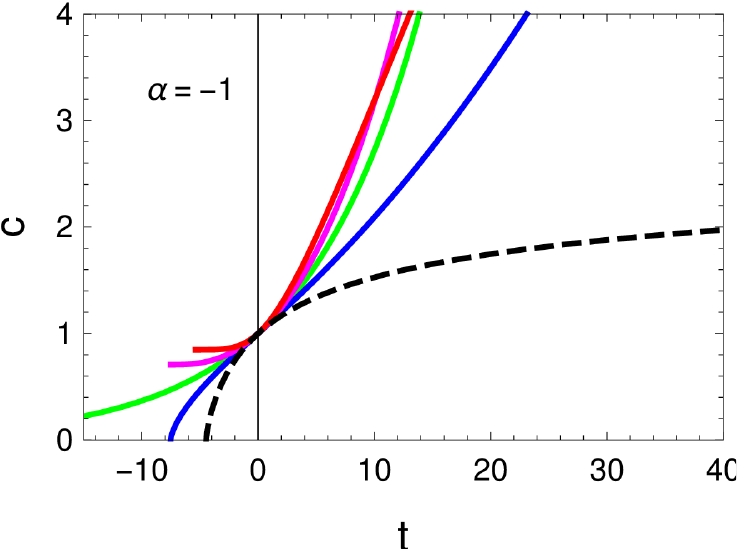}}
\subfigure{\includegraphics[scale=0.44]{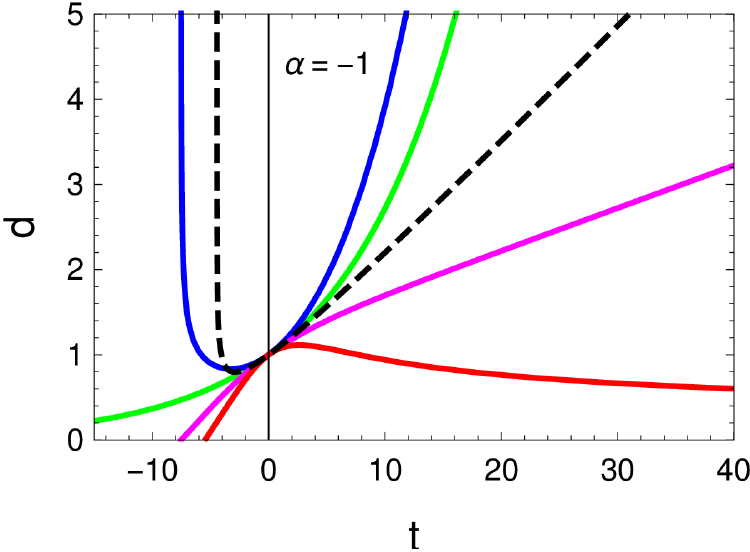}}
\subfigure{\includegraphics[scale=0.145]{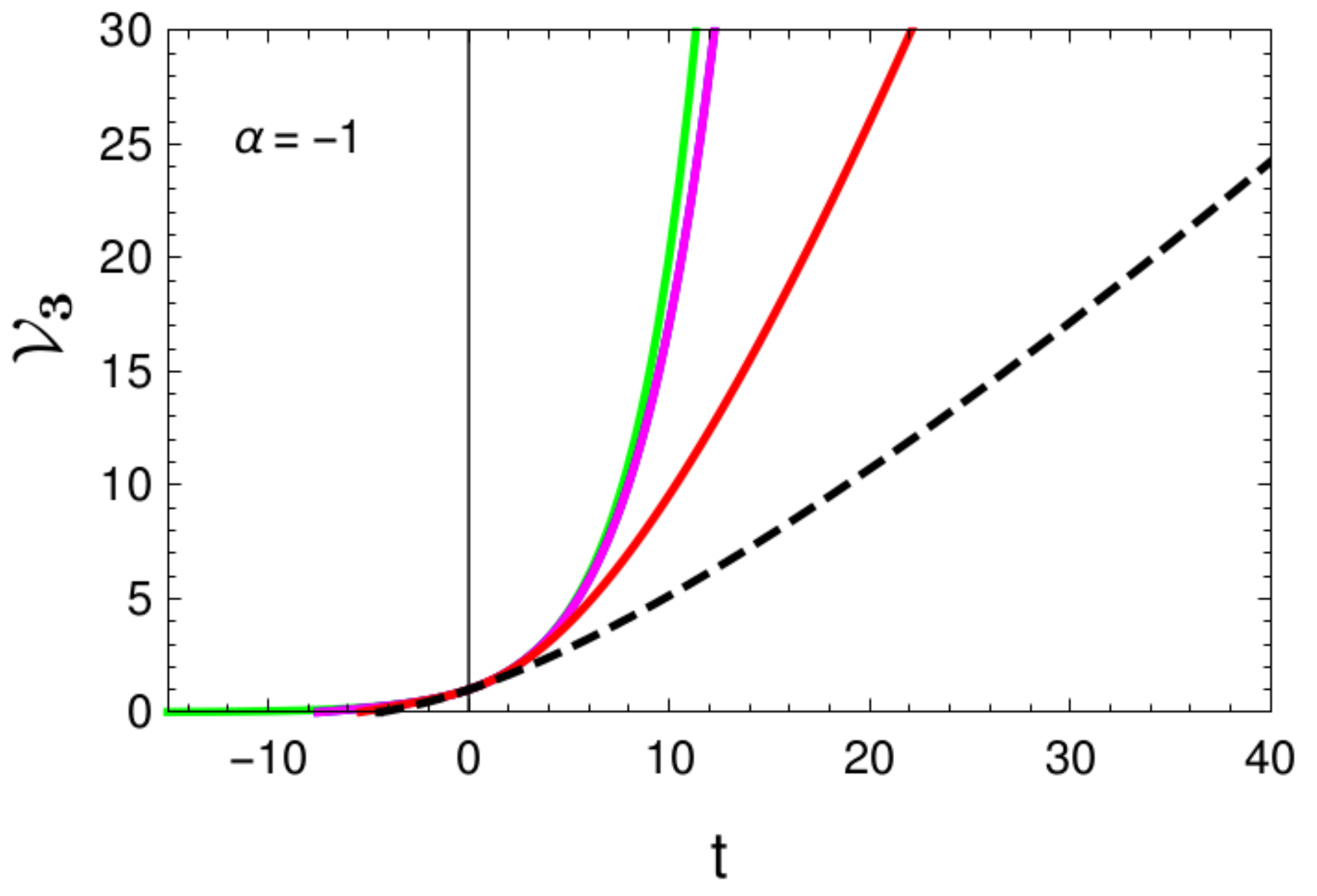}}
\subfigure{\includegraphics[scale=0.48]{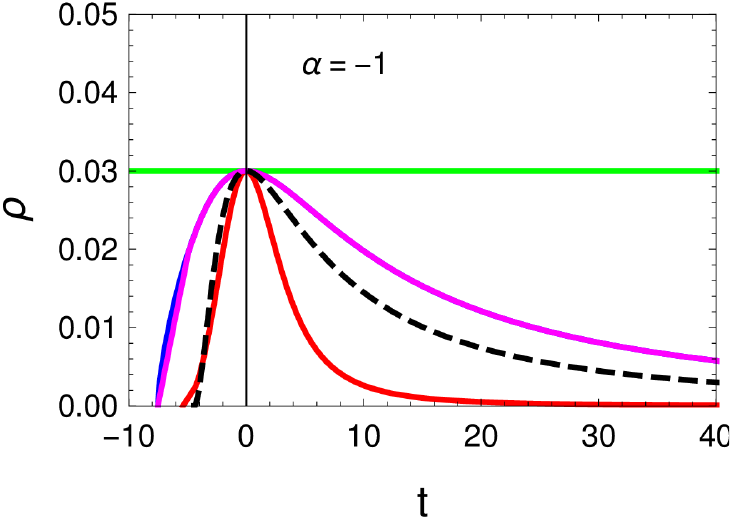}}
\subfigure{\includegraphics[scale=0.48]{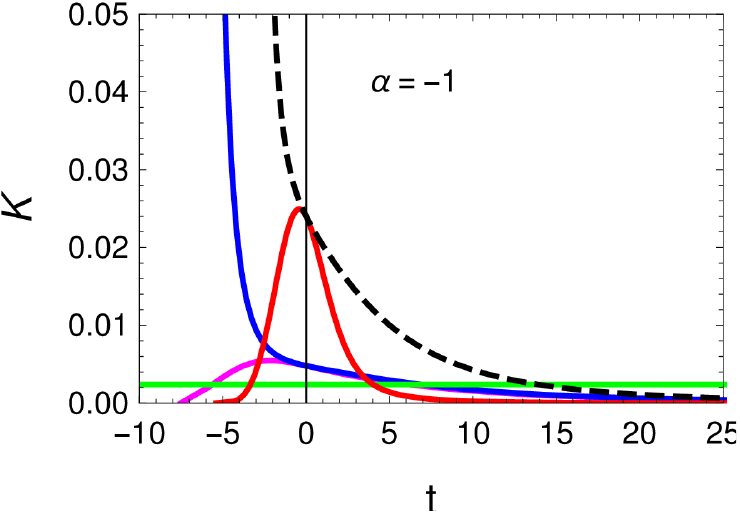}}
\caption{Plots of $c$, $d$, ${\cal V}_3$, $\rho$ and $\mathcal{K}$ 
for $\alpha, \beta = 0, \pm 1/3,\pm 1$.
Plots in a given row are for the same value of $\alpha$. 
Different values of $\beta$ are distinguished by the line color.}
\label{fig1}
\end{figure}

We integrate the $c$-solution in both directions in $t$ from $t=0$.
In Fig.~\ref{fig1}, we plotted the numerical results of the metric functions $c$ and $d$, the three-volume density ${\cal V}_3$, 
the energy density $\rho$, and the Kretschmann scalar ${\cal K}$.
We presented the results for the values of the equation-of-state parameters,
$\alpha, \beta = 0, \pm 1/3,\pm 1$.
The green lines correspond to the usual Friedmann universe ($\beta=0$: stress-free).

%%%%%%%%%%%%%%%%%%%%%%%%%%%%%%%%%%%%%%%%%%%%%%%%%%%%%%%%%%%%%%%%%%%%%%%%%%
%%%%%%%%%%%%%%%%%%%%%%%%%%%%%%%%%%%%%%%%%%%%%%%%%%%%%%%%%%%%%%%%%%%%%%%%%%

\subsection{Metric functions $c$ and $d$, and three-volume density ${\cal V}_3$}

For the usual Friedmann universe ($\beta=0$: green lines in Fig.~\ref{fig1}) 
the metric functions $c$ and $d$ becomes $c=d=a$. 
The functional form of the scale factor $a(t)$ is given similarly by Eq.~\eqref{FRWa} for $\alpha > -1$,
for which the big-bang singularity exists,
\be
a(t) = a_0 (t-t_s)^{2/[3(1+\alpha)]},
\ee
where $t_s$ is the moment of big bang 
(the moment of the three-volume density ${\cal V}_3$ vanishes in our numerical calculations).
For $\alpha = -1$, the scale factor becomes that of de Sitter space and $t_s \to -\infty$.

If $\beta \neq 0$, the metric functions $c$ and $d$ evolve in a different way,
which exhibits an anisotropy (we shall discuss this anisotropy more in detail in Sec.~\ref{sec:aniso}.)
For $\beta >0$, as $t \to t_s$, the metric function $c$ goes to zero and $d$ diverges.
This means that $c$ and $d$ exhibit an expansion and a contraction individually.
For $\beta <0$, as $t \to t_s$, $c$ goes to a nonzero value and $d$ goes to zero.
Both $c$ and $d$ exhibit an expansion.
%For $\beta \neq 0$, the metric function $c$ goes to zero as $t \to t_s$ for $\beta >0$, 
%and goes to nonzero for $\beta <0$,
%while the metric function $d$ diverges as $t \to t_s$ for $\beta >0$, 
%and goes to zero for $\beta <0$.
Although the metric functions $c$ and $d$ behave oppositely as $t \to t_s$,
the three-volume density ${\cal V}_3$ vanishes at $t \to t_s$, 
which exhibits the big bang at the initial moment.
[This is true for all the cases except for de Sitter, $(\alpha,\beta)=(-1,0)$.]
${\cal V}_3$ increases in time monotonically.

\subsection{Energy density $\rho$}

For the usual Friedmann universe ($\beta=0$),
the energy density $\rho$ is given similarly by Eq.~\eqref{FRWrho} for $\alpha > -1$.
It decreases monotonically in $t$ and diverges at the moment of big  bang ($t \to t_s$).
It remains constant in time for the de Sitter space $\alpha = -1$.

For $\beta \neq 0$, however, 
there are some cases which are different from the conventional Friedmann universe.
In Fig.~\ref{fig1}, we observe some cases for which $\rho$ increases in $t$ at the early stage.
In addition, as summarized in Tab.~\ref{tab1}, there are several cases
for which the energy density does not diverge at the initial moment, $\rho (t\to t_s) \nrightarrow \infty$.

\subsection{Kretschmann scalar $\mathcal{K}$}
\label{SecKR}

In the previous subsections, we observed that the three-volume density ${\cal V}_3$ 
vanishes at $t=t_s$ except for the de Sitter space [$(\alpha,\beta)=(-1,0)$]
while the energy density $\rho$ does not necessarily diverge.
Therefore, it is better to evaluate the curvature to examine the character of the spacetime.
The Kretschmann scalar is given by
\begin{equation}
\mathcal{K} = R^{\mu\nu\rho\sigma} R_{\mu\nu\rho\sigma}
=\frac{12A_1^4}{c^{6+2\alpha+4\beta}d^{4\alpha+8\beta}}
+\frac{(3\alpha^2+24\beta^2+2\alpha-8\beta+3)A_1^2 A_2^2}{c^{4+4\alpha-4\beta}d^{2+2\alpha+4\beta}}
+\frac{8(3\beta-1)A_1^3 A_2}{c^{5+3\alpha}d^{1+3\alpha+6\beta}}.
\label{eq:kretschmann}
\end{equation}
We plotted ${\cal K}$ in Fig.~\ref{fig1}, and presented its value at $t=t_s$ in Tab.~\ref{tab1}. 
There are three classes in the relation between $\rho$ an ${\cal K}$ at $t_s$;

(i) for $\rho = \infty$, (A) ${\cal K} = \infty$,

(ii) for $\rho = 0$, (A) ${\cal K} =\infty$  or (B) $0$,

(iii)  for $\rho = {\rm constant}$, (A) ${\cal K} = \infty$  or (C) ${\rm constant}$.

\noindent
Regarding ${\cal V}_3 (t_s)=0$, most of the classes are acceptable except (iii-C)
for which the spacetime is regular while the energy density remains finite with zero volume.
There are possibilities of avoiding the big-bang singularity 
for (ii-B) [$(\alpha,\beta)=(1/3,-1), (0,-1), (-1/3,-1), (-1,-1/3), (-1,-1)$],
and (iii-C) [$(\alpha,\beta)=-1/3,-1/3)$]. 
In particular, it is interesting that for the radiation- and the matter-dominated epochs,
the big-bang singularity can be removed if the shear is $\sigma = -\rho$, i.e.,
$(\alpha,\beta)=(1/3,-1), (0,-1)$.

%Note that, the nature of $\mathcal{K}$ at $t=t_s$ follows $\rho$, 
%except for the cases $(\alpha,\beta)=(\frac{1}{3},1)$, $(0,1)$, 
%$(-\frac{1}{3},1)$, $(-1,1)$, $(-\frac{1}{3},\frac{1}{3})$ and $(-1,\frac{1}{3})$. 
%In these cases, $\mathcal{K}$ diverges at 5$t=t_s$ even though $\rho$ vanishes or is finite.

%%%%%%%%%%%%%%%%%%%%%%%%%%%%%%%%%%%%%%%%%%%%%%%%%%%%%%%%%%%%%%%%%%%%%%%%%%
\begin{table}[h!]
\centering
\caption{Energy density $\rho$ and Kretschmann scalar $\mathcal{K}$ at $t_s$}
 \begin{tabular}{| c || c | c | c | c |} 
 \hline
  $\alpha$ & $\beta$ & $\rho(t_s)$ & $\mathcal{K}(t_s)$ & Nature \\
  \hline\hline
  1 &  $1,\frac{1}{3},0, -\frac{1}{3},-1$ & $\infty$ & $\infty$ & singular \\
   \hline
                &  1 & 0 & $\infty$ & singular \\
  $\frac{1}{3}$ &  $\frac{1}{3},0, -\frac{1}{3}$ & $\infty$ & $\infty$ & singular \\
                &  $-1$ & 0 & 0 & regular \\
   \hline
    &  1 & 0 & $\infty$ & singular \\
  0 &  $\frac{1}{3},0, -\frac{1}{3}$ & $\infty$ & $\infty$ & singular \\
    &  $-1$ & 0 & 0 & regular \\
   \hline
                 &  1 & 0 & $\infty$ & singular \\
                 &  $\frac{1}{3}$ & const. & $\infty$ & singular \\
  $-\frac{1}{3}$ &  0 & $\infty$ & $\infty$ & singular \\
                 &  $-\frac{1}{3}$ & const. & const. & regular \\
                 &  $-1$ & 0 & 0 & regular \\
   \hline
       &  $1,\frac{1}{3}$ & 0 & $\infty$ & singular \\
  $-1$ &  0 & const. & const. & regular \\
       &  $-\frac{1}{3},-1$ & 0 & 0 & regular \\
 \hline
 \end{tabular}
\label{tab1}
\end{table}
%%%%%%%%%%%%%%%%%%%%%%%%%%%%%%%%%%%%%%%%%%%%%%%%%%%%%%%%%%%%%%%%%%%%%%%%%%

\subsection{$\rho(t_s)$ \& $\mathcal{K}(t_s)$}
Let us discuss the initial singularity more in detail.
We investigate the solutions in the limit of $t\to t_s$ 
in order to inspect the behaviour of $\rho(t_s)$ and $\mathcal{K}(t_s)$. 
At $t=0$, we imposed the initial conditions, $c(0)=d(0)=1$ and $\dot{c}(0)=\dot{d}(0)$. 
They give $A_2=3A_1$ from Eq.~\eqref{eq:cd_gen} and 
$A_3=4\beta/(1-\alpha+2\beta)$ from Eq.~\eqref{eq:d_gen}. 
Considering the signatures of the terms in the integrand, 
the $c$-solution for Class G in Eq.~\eqref{eq:t_gen} is classified into four cases,
\begin{equation}
t=\left\{
  \begin{array}{lr}
  \frac{1}{A_1}\int_{1}^c \sqrt{c} 
  \left[\frac{1-\alpha-2\beta}{1-\alpha+2\beta}
  \left( c^{3(1-\alpha+2\beta)/2}
  -c_{*}^{3(1-\alpha+2\beta)/2} \right)\right]^\frac{\alpha+2\beta}{1-\alpha-2\beta} dc 
  & (\mbox{G1: } \beta<0\; \& \; 1-\alpha+2\beta > 0) \\
  \frac{1}{A_1}\int_{1}^c \sqrt{c} 
  \left[\frac{1-\alpha-2\beta}{|1-\alpha+2\beta|}
  \left(-c^{-3|1-\alpha+2\beta|/2}
  +c_{*}^{-3|1-\alpha+2\beta|/2} \right)\right]^\frac{\alpha+2\beta}{1-\alpha-2\beta} dc 
  & (\mbox{G2: } \beta<0\; \& \; 1-\alpha+2\beta < 0) \\
  \frac{1}{A_1}\int_{1}^c \sqrt{c} 
  \left[\frac{1-\alpha-2\beta}{1-\alpha+2\beta}
  \left( c^{3(1-\alpha+2\beta)/2}
  +c_{*}^{3(1-\alpha+2\beta)/2} \right)\right]^\frac{\alpha+2\beta}{1-\alpha-2\beta} dc 
  & (\mbox{G3: } \beta>0\; \& \; 1-\alpha-2\beta > 0) \\
  \frac{1}{A_1}\int_{1}^c \sqrt{c} 
  \left[\frac{|1-\alpha-2\beta|}{1-\alpha+2\beta}
  \left( -c^{3(1-\alpha+2\beta)/2}
  +c_{*}^{3(1-\alpha+2\beta)/2} \right)\right]^\frac{\alpha+2\beta}{1-\alpha-2\beta} dc 
  & (\mbox{G4: } \beta>0\; \& \; 1-\alpha-2\beta < 0) 
  \end{array}
\right.,
\label{c_eq_4}
\end{equation}
where $c_{*}=(4|\beta|/|1-\alpha-2\beta|)^{2/[3(1-\alpha+2\beta)]}$. 
Note that for $-1\leq \alpha\leq 1$, 
if $\beta>0$, we have $1-\alpha+2\beta >0$, 
and if $\beta<0$, we have $1-\alpha-2\beta >0$. 
As $t\to t_s$, $c$ approaches to its minimum value; 
$c\to c_{*}$ for G1 and G2, and $c\to 0$ for G3 and G4. 
We split the integration range into two parts as
\begin{equation}
\int_1^c[\cdots]dc = \left( \int_1^{c_{s}} + \int_{c_{s}}^c \right)[\cdots]dc, 
\quad \mbox{where }
\int_1^{c_{s}}[\cdots]dc\equiv t_s 
\mbox{ and } c_s=\left\{
  \begin{array}{l}
  \mbox{$c_*$ for G1 and G2} \\
  \mbox{$0$ for G3 and G4}
  \end{array}
  \right..
\end{equation}
%Since $c\to c_s$ as $t\to t_s$, 
%the first part in the above integration vanishes in this limit, 
%and the second part gives $t_s$. 
Therefore, Eq.~\eqref{c_eq_4} can be rewritten as
\begin{equation}
t-t_s=\left\{
  \begin{array}{lr}
  \frac{1}{A_1}\int_{c_{*}}^c \sqrt{c} 
  \left[\frac{1-\alpha-2\beta}{1-\alpha+2\beta}
  \left( c^{3(1-\alpha+2\beta)/2}
  -c_{*}^{3(1-\alpha+2\beta)/2} \right)\right]^\frac{\alpha+2\beta}{1-\alpha-2\beta} dc 
  & (\mbox{G1}) \\
  \frac{1}{A_1}\int_{c_{*}}^c \sqrt{c} 
  \left[\frac{1-\alpha-2\beta}{|1-\alpha+2\beta|}
  \left(-c^{-3|1-\alpha+2\beta|/2}
  +c_{*}^{-3|1-\alpha+2\beta|/2} \right)\right]^\frac{\alpha+2\beta}{1-\alpha-2\beta} dc 
  & (\mbox{G2}) \\
  \frac{1}{A_1}\int_{0}^c \sqrt{c} 
  \left[\frac{1-\alpha-2\beta}{1-\alpha+2\beta}
  \left( c^{3(1-\alpha+2\beta)/2}
  +c_{*}^{3(1-\alpha+2\beta)/2} \right)\right]^\frac{\alpha+2\beta}{1-\alpha-2\beta} dc 
  & (\mbox{G3}) \\
  \frac{1}{A_1}\int_{0}^c \sqrt{c} 
  \left[\frac{|1-\alpha-2\beta|}{1-\alpha+2\beta}
  \left( -c^{3(1-\alpha+2\beta)/2}
  +c_{*}^{3(1-\alpha+2\beta)/2} \right)\right]^\frac{\alpha+2\beta}{1-\alpha-2\beta} dc 
  & (\mbox{G4}) 
  \end{array}
\right..
\end{equation}
We perform the Taylor expansion for the term in the parentheses of the integrand 
about $c=c_{*}$ for G1 and G2 and keep the leading order. For G3 and G4, we set $c=0$ in the parentheses and get the leading order of the integrand.
Keeping the leading order, the integration gives, in the limit of $t \to t_s$,
\begin{equation}
c(t)\sim \left\{
  \begin{array}{ll}
  c_{*}+{\rm const.}\times (t-t_s)^{(1-\alpha-2\beta)} & (\mbox{G1 and G2}) \\
  (t-t_s)^{2/3} 
  & (\mbox{G3 and G4})
  \end{array}
\right..
\end{equation}
Using this result of $c$,
we obtain $d$ from Eq.~\eqref{eq:d_gen},
\begin{equation}
d\sim \left\{
  \begin{array}{ll}
  t-t_s  & (\mbox{G1 and G2}) \\
  (t-t_s)^{-1/3}  & (\mbox{G3 and G4})
  \end{array}
\right..
\end{equation}
As a result, the three-volume density becomes 
\begin{equation}
{\cal V}_3 \sim t-t_s \quad\mbox{(G1-G4)}, 
\end{equation}
the energy density becomes
\begin{equation}
\rho \sim \left\{
  \begin{array}{ll}
  \frac{1}{(t-t_s)^{1+\alpha+2\beta}} 
  & (\mbox{G1 and G2}) \\
  \frac{1}{(t-t_s)^{1+\alpha-2\beta}} 
  & (\mbox{G3 and G4}) 
  \end{array}
\right.,
\end{equation}
and the Kretschmann scalar becomes
\begin{equation}
\mathcal{K} \sim \left\{
  \begin{array}{ll}
  \frac{1}{(t-t_s)^{2(1+\alpha+2\beta)}} 
  & (\mbox{G1 and G2}) \\
  \frac{1}{(t-t_s)^4} 
  & (\mbox{G3 and G4}) 
  \end{array}
\right..
\end{equation}

\begin{figure}[]
\centering
\includegraphics[scale=0.25]{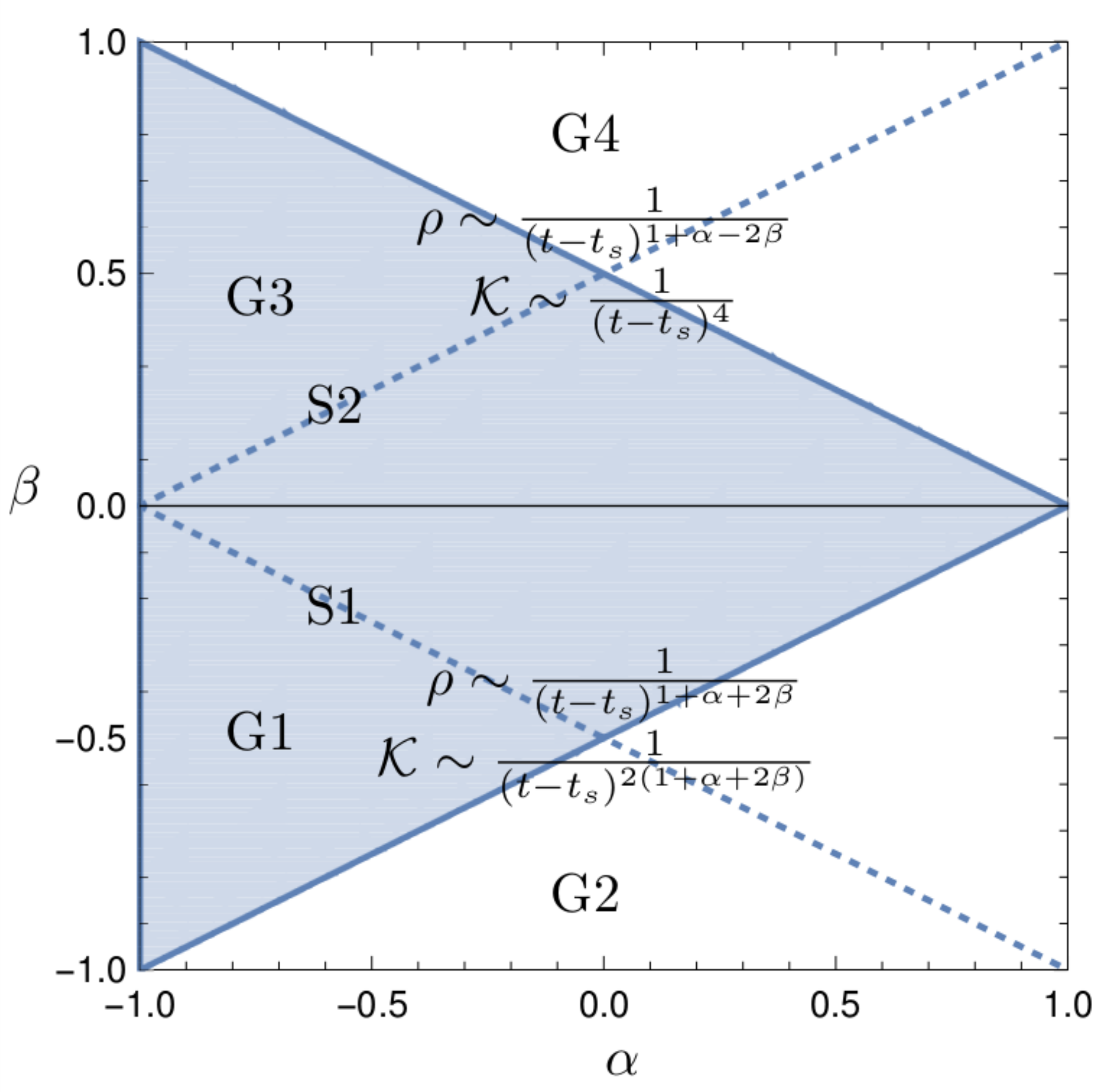}
\caption{Diagram for $\rho$ and $\mathcal{K}$ at $t\to t_s$
in $\alpha$-$\beta$ plane.
Class G is split into four triangular regions by G1-G4.
The shaded region ($1-\alpha-2|\beta| > 0$) consists  of G1 and G3 divided by $\beta=0$,
and the unshaded region ($1-\alpha-2|\beta| < 0$) consists of G2 and G4. 
The dotted lines S1 and S2 represent $1+\alpha+2\beta=0$ and $1+\alpha-2\beta=0$, respectively. 
Note that the initial singularity can be removed on S1 and in the region below.}
\label{fig2}
\end{figure}

The behaviour of $\rho(t_s)$ and ${\cal K}(t_s)$ can be characterized by the regions 
in the $\alpha$-$\beta$ plane in Fig.~\ref{fig2}. Note that the line S1 ($1+\alpha+2\beta=0$) divides each of the regions G1 and G2 into two parts, and that S2 ($1+\alpha-2\beta=0$) does similarly for G3 and G4.

1) In the region below S1, $\rho(t_s)={\cal K}(t_s)=0$ (e.g., ii-B in Sec.~\ref{SecKR}); the solution is non-singular. This is a part of G1 and G2 (see Fig.~\ref{fig2}).

2) On S1, $\rho(t_s)$ and ${\cal K}(t_s)$ are finite (iii-C); the solution is non-singular.

3) In the region between S1 and S2, $\rho(t_s)$ and ${\cal K}(t_s)$ diverge (i-A); the solution is singular. This is a part of G1-G4.

4) In the region above S2, $\rho(t_s)=0$ and ${\cal K}(t_s)$ diverges (ii-A); the solution is singular. This is a part of G3 and G4.

5) On S2, $\rho(t_s)$ is finite and ${\cal K}(t_s)$ diverges (iii-A); the solution is singular.

\noindent
Therefore, the initial singularity is removed only in the region on and below S1.

Let us discuss the energy conditions. The standard energy conditions \citep{poisson} for the diagonal energy-momentum tensor in Eq. \eqref{EMTdiag} are as following;\\
(a) Weak Energy Condition (WEC):
\begin{equation}
\rho\geq 0, \;\; \rho+p-\sigma>0, \;\; \rho+p+2\sigma>0 \quad \Rightarrow \quad \rho\geq 0, \;\; 1+\alpha-\beta>0, \;\; 1+\alpha+2\beta>0.
\nonumber
\end{equation}
(b) Null Energy Condition (NEC):
\begin{equation}
\rho+p-\sigma\geq 0, \;\; \rho+p+2\sigma\geq 0 \quad \Rightarrow \quad 1+\alpha-\beta\geq 0, \;\; 1+\alpha+2\beta\geq 0.
\nonumber
\end{equation}
(c) Strong Energy Condition (SEC):
\begin{equation}
\rho+p-\sigma\geq 0, \;\; \rho+p+2\sigma\geq 0, \;\; \rho+3p\geq 0 \quad \Rightarrow \quad 1+\alpha-\beta\geq 0, \;\; 1+\alpha+2\beta\geq 0, \;\; 1+3\alpha\geq 0.
\nonumber
\end{equation}
(d) Dominant Energy Condition (DEC):
\begin{equation}
\rho\geq 0, \;\; \rho\geq |p-\sigma|, \;\; \rho\geq |p+2\sigma| \quad \Rightarrow \quad \rho\geq 0, \;\; 1-|\alpha-\beta|\geq 0, \;\; 1-|\alpha+2\beta|\geq 0.
\nonumber
\end{equation}
The energy conditions are written in $\alpha$ and $\beta$ using the equations of state in Eq. \eqref{eoss}. In the region below and on S1, we have $1+\alpha+2\beta\leq 0$. Therefore, ``all the energy conditions" are violated in the region below S1. Only the WEC is violated on S1. The condition for the energy-condition violation coincides with the condition for the removal of the singularity of our work.

Belinski and Khalatnikov \citep{VC1-1} studied anisotropic Bianchi type-I cosmology with bulk and shear viscosities qualitatively through dynamical-system studies. In the asymptotic limits of both small and large $\rho$, they considered $\eta\propto \rho^A$, $A$ being a constant. They showed, however, that the cosmological singularity is not removed. In a slightly different context, Medina et. al. \citep{VC7} suggested that the universe can bypass the initial singularity in the Einstein-Cartan theory.

%%%%%%%%%%%%%%%%%%%%%%%%%%%%%%%%%%%%%%%%%%%%%%%%%%%%%%%%%%%%%%%%%%%%%%%%%%
%%%%%%%%%%%%%%%%%%%%%%%%%%%%%%%%%%%%%%%%%%%%%%%%%%%%%%%%%%%%%%%%%%%%%%%%%%

\newpage
\section{Late time behaviour}
\label{sec:latetime}

\begin{figure}[]
\centering
\includegraphics[scale=0.25]{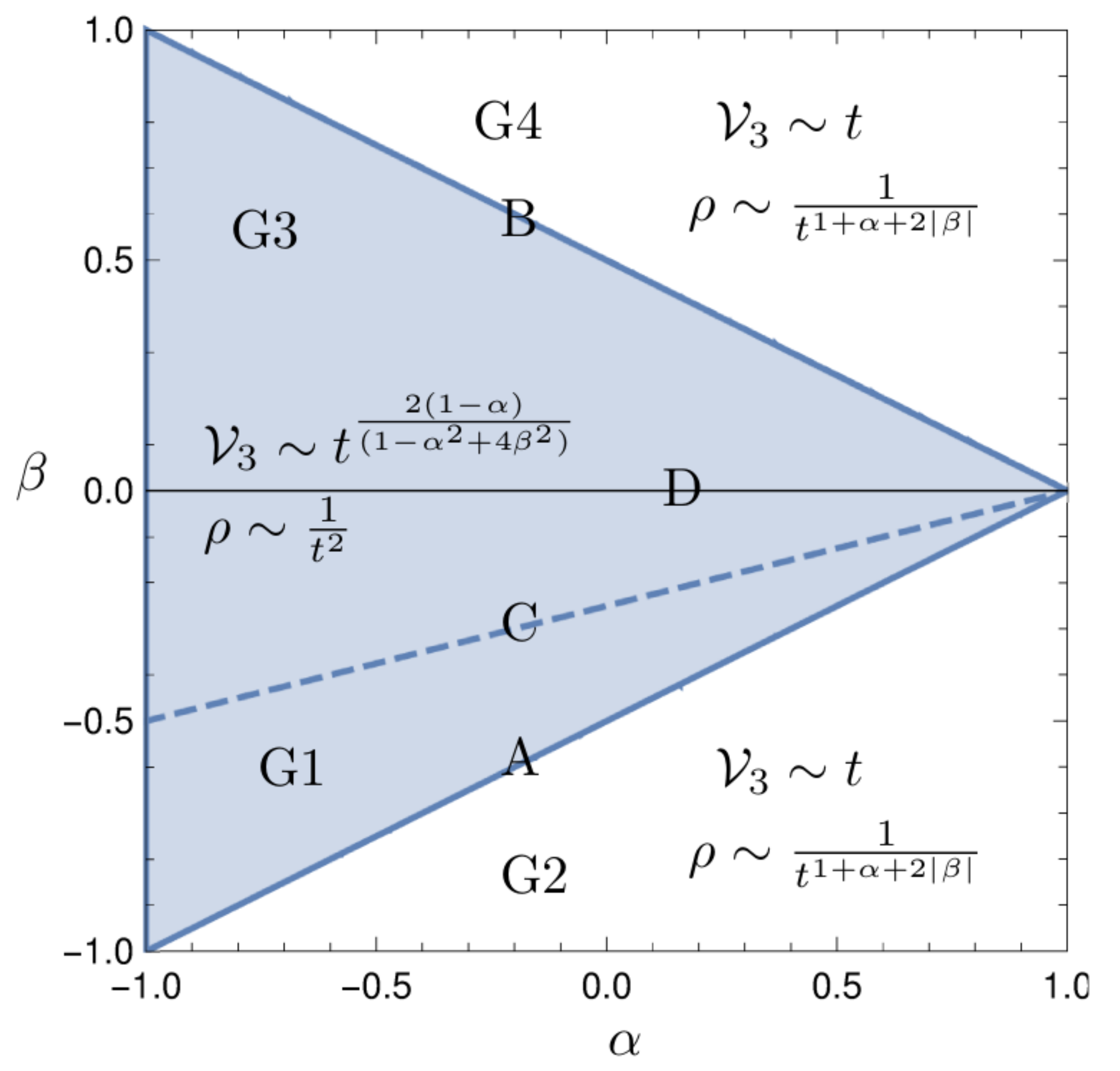}
\caption{Diagram for ${\cal V}_3$ and $\rho$ at large $t$
in $\alpha$-$\beta$ plane.
The lower border between the shaded and the unshaded regions
corresponds to Class A,
and the upper border to Class B.
The dashed line in G1 corresponds to Class C,
and the solid horizontal line between G1 and G3 to Class D ($\beta=0$: FRW). 
Considering the power in $t$,
$\rho$ evolves in the same manner with that of FRW in the shaded region, 
and decreases faster than that of FRW in the unshaded region.
${\cal V}_3$ increases slower than that of FRW in all regions.
}
\label{fig3}
\end{figure}

In this section, we investigate the late-time behaviour of the solutions at large $t$. 
We focus on different classes separately. 
We summarized the results in Fig.~\ref{fig3}.

\vspace{12pt}
\noindent
\underline{{\bf Class G}: general class}
\vspace{12pt}

As $t\to \infty$, we expect that the integration in Eq.~\eqref{c_eq_4} diverges. 
For Class G1 and G2, 
we observed from our calculations 
that the integration diverges as $c\to\infty$ and becomes finite as $c\to c_{*}$. 
For the class G3,
the integration diverges also as $c\to\infty$.
For the class G4, however,
the integration diverges as $c\to c_{*}$.

After carrying out integrations, 
we find $c$ at large $t$,
\begin{equation}\label{clate}
c(t)\sim \left\{
  \begin{array}{ll}
  t^{2(1-\alpha-2\beta)/[3(1-\alpha^2+4\beta^2)]} 
  & (\mbox{G1 and G3}) \\
  t^{2/3} 
  & (\mbox{G2}) \\
  c_{*}-{\rm const.}\times t^{-|1-\alpha-2\beta|} 
  & (\mbox{G4}) 
  \end{array}
\right..
\end{equation}
Using the results of $c$ in the limit of $c\to \infty$, or $c\to c_{*}$,
we obtain $d$ from Eq.~\eqref{eq:d_gen} as a function of $c$ at large $t$,
\begin{equation}\label{dlate}
d\sim \left\{
  \begin{array}{ll}
  c^{(1-\alpha+4\beta)/(1-\alpha-2\beta)} 
   & (\mbox{G1 and G3}) \\
  c^{-1/2} 
   & (\mbox{G2}) \\
  (c_{*}-c)^{-1/|1-\alpha-2\beta|}
  & (\mbox{G4}) 
  \end{array}
\right..
\end{equation}
The three-volume density becomes
\begin{equation}
{\cal V}_3 \sim \left\{
  \begin{array}{ll}
  t^{2(1-\alpha)/(1-\alpha^2+4\beta^2)} 
  & (\mbox{G1 and G3}) \\
  t 
  & (\mbox{G2 and G4}) 
  \end{array}
\right..
\label{GV3LT}
\end{equation}
The energy density becomes
\begin{equation}
\rho\sim \left\{
  \begin{array}{ll}
  \frac{1}{t^2} 
  & (\mbox{G1 and G3}) \\
  \frac{1}{t^{1+\alpha+2|\beta|}}
  & (\mbox{G2 and G4}) 
  \end{array}
\right..
\label{Grho3LT}
\end{equation}
Note that ${\cal V}_3$ diverges and $\rho$ vanishes as $t\to \infty$.
Class G1-G4 correspond to each rectangular triangle in Fig.~\ref{fig3}.  

%%%%%%%%%%%%%%%%%%%%%%%%%%%%%%%%%%%%%%%%%%%%%%%%%%%%%%%%%%%%%%%%%%%%%%%%%%
%%%%%%%%%%%%%%%%%%%%%%%%%%%%%%%%%%%%%%%%%%%%%%%%%%%%%%%%%%%%%%%%%%%%%%%%%%

\vspace{12pt}
\noindent
\underline{{\bf Class C}: $1-\alpha+4\beta=0$ ($\alpha\neq 1$ and $\beta\neq 0$)}
\vspace{12pt}

Imposing the same initial conditions with the Class G at $t=0$,
we have the same relation $A_2=3A_1$ from Eq.~\eqref{eq:cd_SC3},
but have $A_{3C}=2$ from Eq.~\eqref{eq:d_SC3}.
Then Eq.~\eqref{eq:t_SC3} becomes
\begin{equation}\label{eq:t_SC3_2}
t=\frac{1}{A_1} \int_{1}^c c^{(3\alpha+1)/4} 
\left[3- 2 c^{-3(1-\alpha)/4} \right]^{\frac{3\alpha-1}{3(1-\alpha)}} dc.
\end{equation}
This integration diverges as $c\to\infty$.
Then from Eqs.~\eqref{eq:t_SC3_2} and \eqref{eq:d_SC3}, we have 
\be
c \sim t^{4/(3\alpha+5)},\qquad
d= {\rm const.}
\ee
The three-volume density and the energy density at large $t$ are given by
\begin{equation}
{\cal V}_3 \sim t^{8/(3\alpha+5)}, \qquad \rho\sim \frac{1}{t^2}.
\end{equation}
Class C corresponds to the dashed line in Fig~\ref{fig3}. Note that this class is contained in G1.

\vspace{12pt}
\noindent
\underline{{\bf Class D}: $\beta=0$}
\vspace{12pt}

{\bf (i) FRW universe}: 
For this class, the metric is diagonal, 
and $c$ and $d$ were obtained in an exact form, Eq.~\eqref{eq:c_SC4}. 
At large $t$, we have
\begin{equation}
c=d\sim \left\{
  \begin{array}{lr}
    e^{A_1 t} &(\alpha =-1) \\
  t^{2/[3(1+\alpha)]} &(\alpha \neq -1)
  \end{array}
\right..
\end{equation}
The three-volume density and the energy density at large $t$ are given by
\begin{equation}
{\cal V}_3 \sim \left\{
  \begin{array}{lr}
    e^{3A_1 t} & (\alpha =-1)\\
  t^{2/(1+\alpha)} & (\alpha \neq -1)
  \end{array}
\right., \qquad 
\rho\sim \left\{
  \begin{array}{lr}
    {\rm const.} & (\alpha =-1) \\
  \frac{1}{t^2} & (\alpha \neq -1)
  \end{array}
\right..
\end{equation}
Class D corresponds to the horizontal axis in Fig.~\ref{fig3}.

{\bf (ii) Non-FRW universe}: 
For this class, the metric is non-diagonal, 
and $c$ and $d$ are given by Eqs. \eqref{eq:d_SC4} and \eqref{eq:t_SC4}.
However, we find that the late-time behaviours of ${\cal V}_3$ and $\rho$ are the same as Class D-(i).

\vspace{12pt}
\noindent
\underline{{\bf Class A} ($1-\alpha+2\beta=0$) and {\bf Class B} ($1-\alpha-2\beta=0$)}
\vspace{12pt}

For Class A and  B, it is difficult to obtain the solution at large $t$
from directly integrating Eqs.~\eqref{eq:t_SC1} and \eqref{eq:t_SC2}.
However, since Class A is the border of G1 and G2,
and Class B is the border of G3 and G4 
(i.e., the border lines between the shaded and unshaded regions in Fig.~\ref{fig3}),
we can deduce ${\cal V}_3$ and $\rho$.
The conditions for Class A and B can be written as
$1-\alpha-2|\beta|=0$ which gives $2|\beta|=1-\alpha$. 
From Eqs.~\eqref{GV3LT} and \eqref{Grho3LT}, 
we observe that the functional forms of ${\cal V}_3$ and $\rho$ converge at the  borders,
\begin{equation}
{\cal V}_3 \sim t, \qquad \rho\sim \frac{1}{t^2}.
\end{equation}

\vspace{12pt}
The behaviours of ${\cal  V}_3$ and $\rho$ at large $t$ are summarized in the diagram in Fig.~\ref{fig3}. 
Consider the power-law dependence of $\rho$ and ${\cal  V}_3$ on $t$. 
In the shaded region (regions G1 and G3),
$\rho$ evolves in the same manner with that of FRW (Class D),
while ${\cal  V}_3$ exhibits a slower expansion than that of FRW.
In the unshaded region (regions G2 and G4),
$\rho$ exhibits a more rapid decrease than that of FRW,
while ${\cal  V}_3$ evolves linearly in $t$ 
implying a slower expansion than that of FRW.
Therefore, with off-diagonal stress,
the expansion of ${\cal V}_3$ is slower, 
and $\rho$ decreases more rapidly (at most equally) compared with the FRW universe.

The expansion of ${\cal V}_3$ can be explained by the Raychaudhuri equation which is
\begin{equation}
\dot{\Theta}+\frac{1}{3}\Theta^2+\sigma^2-\omega^2+R_{\mu\nu}u^\mu u^\nu=0.
\end{equation}
We note from Sec. \ref{sec:model} that $\Theta=2\dot{c}/c+\dot{d}/d=\dot{{\cal V}}_3/{\cal V}_3$ 
and $\sigma^2=\sigma_{\mu\nu}\sigma^{\mu\nu}=2(\dot{c}/c-\dot{d}/d)^2/3$. 
For the FRW universe($\beta=0$), $\sigma^2=0$ as $c=d$ always. 
However, the shear term is nonzero for $\beta\neq 0$. 
Therefore, the positive contribution from $\sigma^2$ makes the expansion $\Theta$ slower.

The behaviour of $\rho$ can be explained by the conservation equation,
\begin{equation}
\dot{\rho}+\left(2\frac{\dot{c}}{c}+\frac{\dot{d}}{d}\right)(\rho+p)-2\left(\frac{\dot{c}}{c}-\frac{\dot{d}}{d}\right)\sigma=0.
\end{equation}
At late times, the off-diagonal stress term behaves as
\begin{equation}
-2\left(\frac{\dot{c}}{c}-\frac{\dot{d}}{d}\right)\sigma\sim \left\{
  \begin{array}{ll}
  \frac{|\beta|}{t^3} 
  & (\mbox{G1 and G3}) \\
  \frac{|\beta|}{t^{2+\alpha+2|\beta|}}
  & (\mbox{G2 and G4}) 
  \end{array}
\right..
\end{equation}
This term contributes negatively to the evolution of $\rho$, 
and make $\rho$ decreases faster than in the FRW universe.
As a whole, the energy density leaks to the shear stress, so drops faster.
This makes the expansion of the Universe slower.

\section{Anisotropy}
\label{sec:aniso}

\begin{figure}[]
\centering
\subfigure{\includegraphics[scale=0.44]{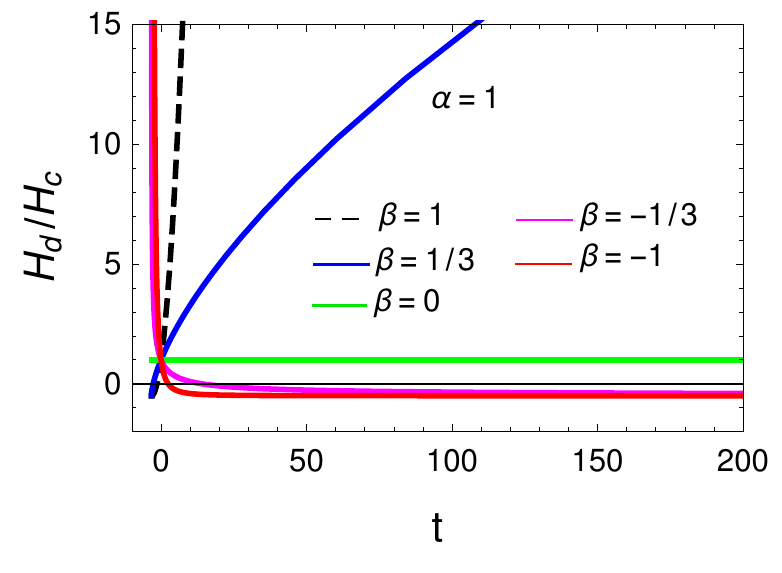}}
\subfigure{\includegraphics[scale=0.44]{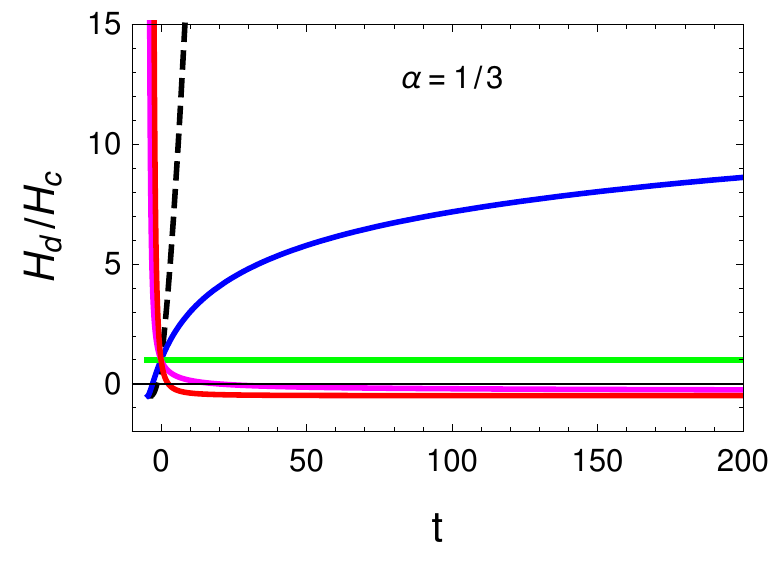}}
\subfigure{\includegraphics[scale=0.44]{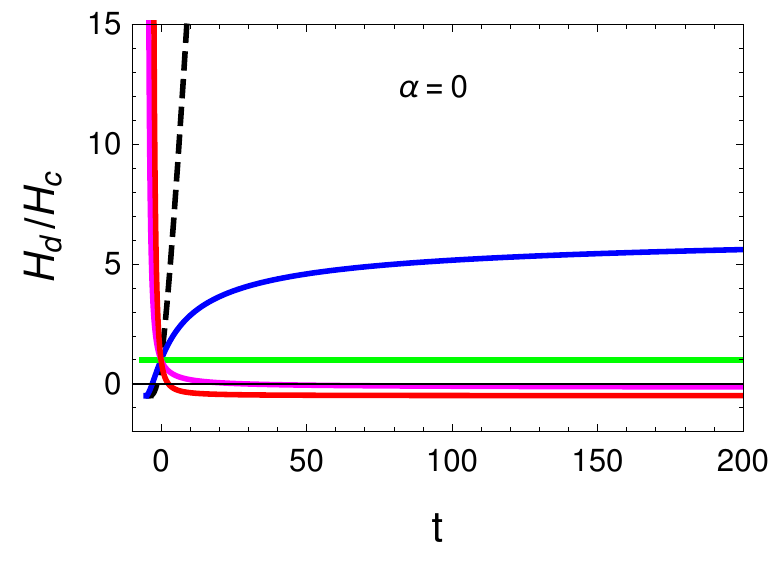}}
\subfigure{\includegraphics[scale=0.44]{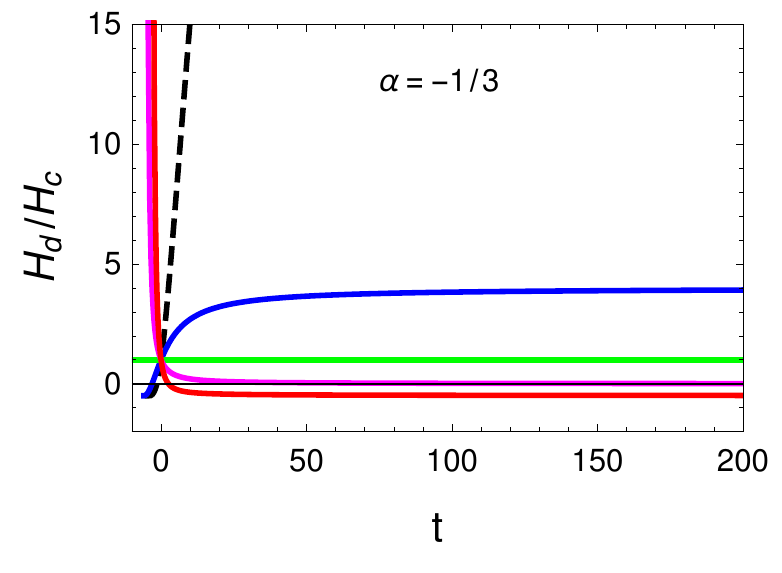}}
\subfigure{\includegraphics[scale=0.44]{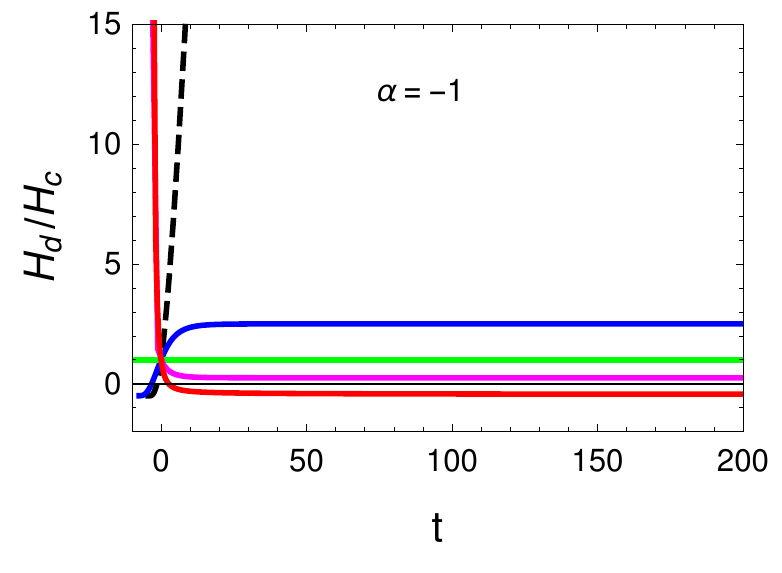}}
\subfigure{\includegraphics[scale=0.44]{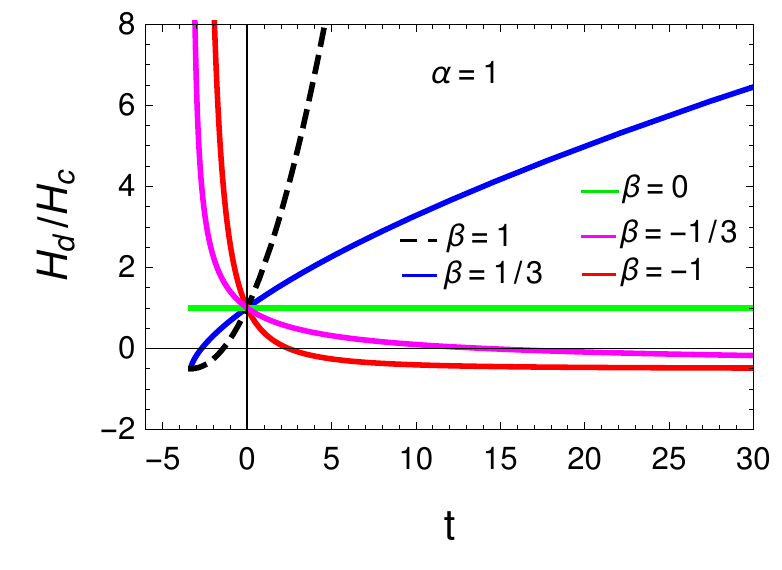}}
\subfigure{\includegraphics[scale=0.44]{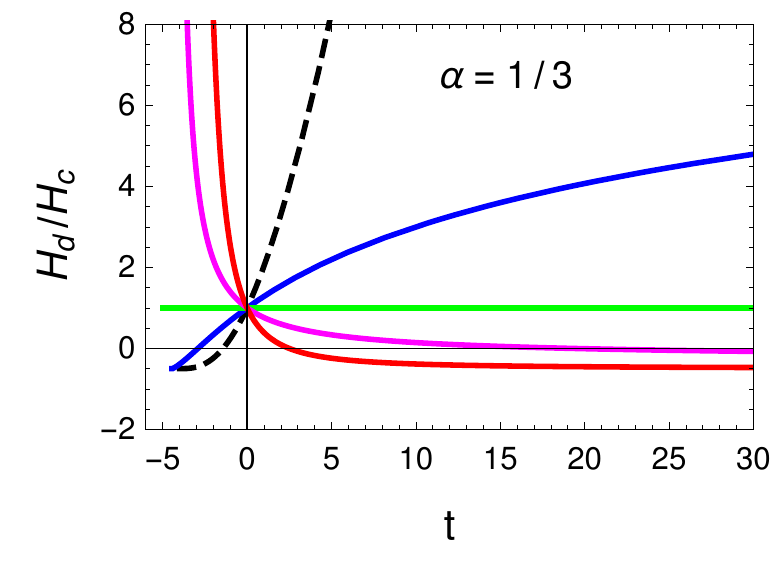}}
\subfigure{\includegraphics[scale=0.44]{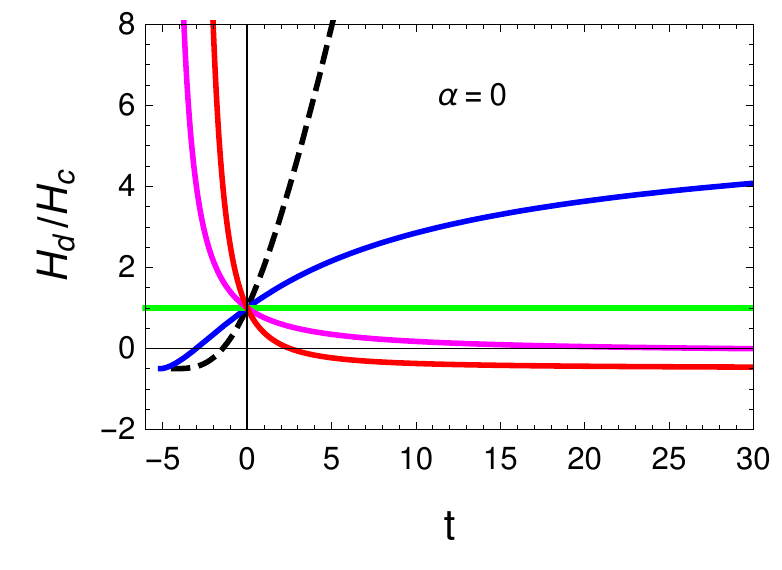}}
\subfigure{\includegraphics[scale=0.44]{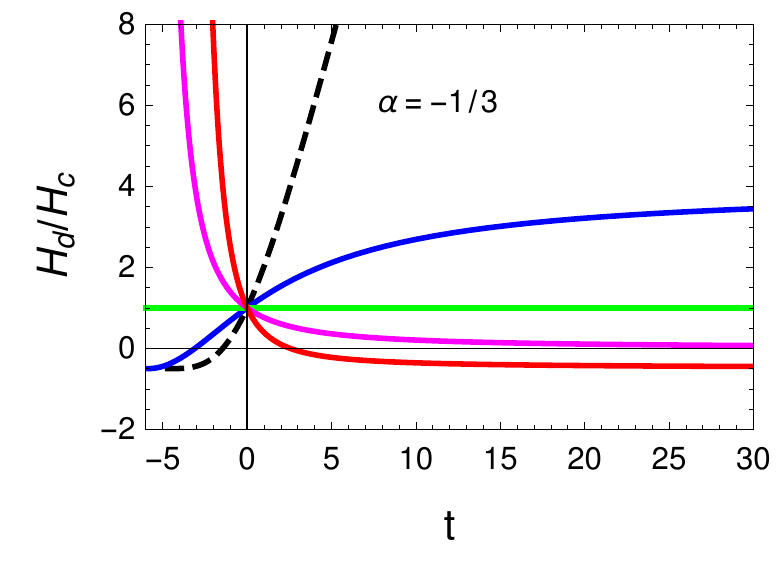}}
\subfigure{\includegraphics[scale=0.44]{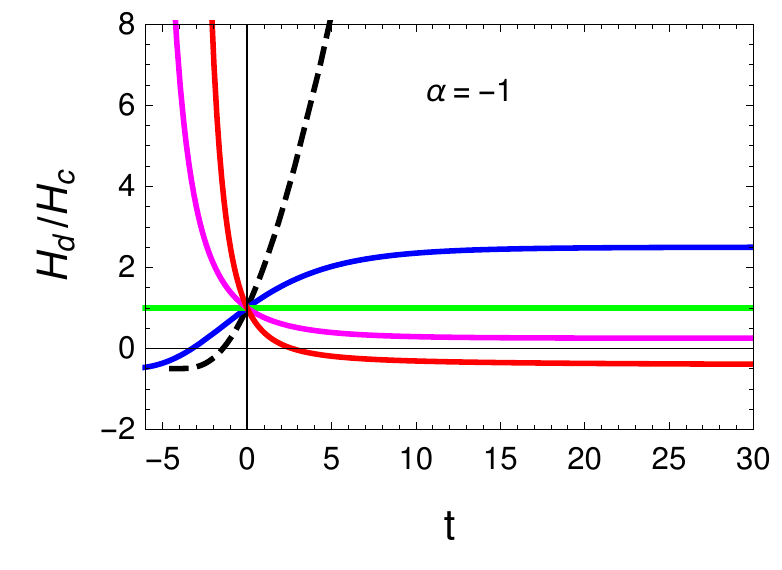}}
\caption{Plots of $H_d/H_c$. (Lower panel is for zoom-in plots.)
At late times, the anisotropy approaches a constant except for G4.
The anisotropy is large unless $\beta$ is small.
The negative value at late times indicates the contraction of $d$,
which occurs in the region below the line C in the diagram of Fig.~\ref{fig3}.}
\label{fig4}
\end{figure}

In this section, we discuss the anisotropy of our model.
Since our model corresponds to the Bianchi type VII as seen from the metric \eqref{metricBianchi}, 
we define the anisotropy $H_d/H_c \equiv (\dot{d}/d)/(\dot{c}/c)$.
It is plotted in Fig.~\ref{fig4}.
Depending on the values of parameters, 
the initial anisotropy decreases from a large value,
or increases from a small value.
At late times, however, the anisotropy approaches a constant value except for the class G4.
The asymptotic value of the anisotropy can be evaluated at late times from Eqs.~\eqref{clate} and \eqref{dlate},
\begin{equation}\label{Anisolate}
\frac{H_d}{H_c} 
\sim \left\{
  \begin{array}{ll}
  \frac{1-\alpha+4\beta}{1-\alpha-2\beta} 
   & (\mbox{G1 and G3}) \\
  -\frac{1}{2}
   & (\mbox{G2}) \\
  \frac{1}{|1-\alpha-2\beta|} \left( -1 +\frac{c_*}{\rm const.} t^{|1-\alpha-2\beta|}\right)
  & (\mbox{G4}) 
  \end{array}
\right..
\end{equation}
Since the anisotropy is due to the off-diagonal stress, 
its magnitude is not negligible unless $\beta$ is small.
The situation is different from the rapid decay of anisotropy with a cosmological constant 
in the background \cite{Wald:1983ky}.

At late times, one can check from Eqs.~\eqref{clate} and \eqref{dlate}
that $H_c>0$ in all the regions of the diagram in Fig.~\ref{fig3}
and $H_d>0$ only in the region above the line C.
It means $d$ exhibits a contraction at late times in the region below the line C.

\begin{figure}[]
\centering
\subfigure{\includegraphics[scale=0.75]{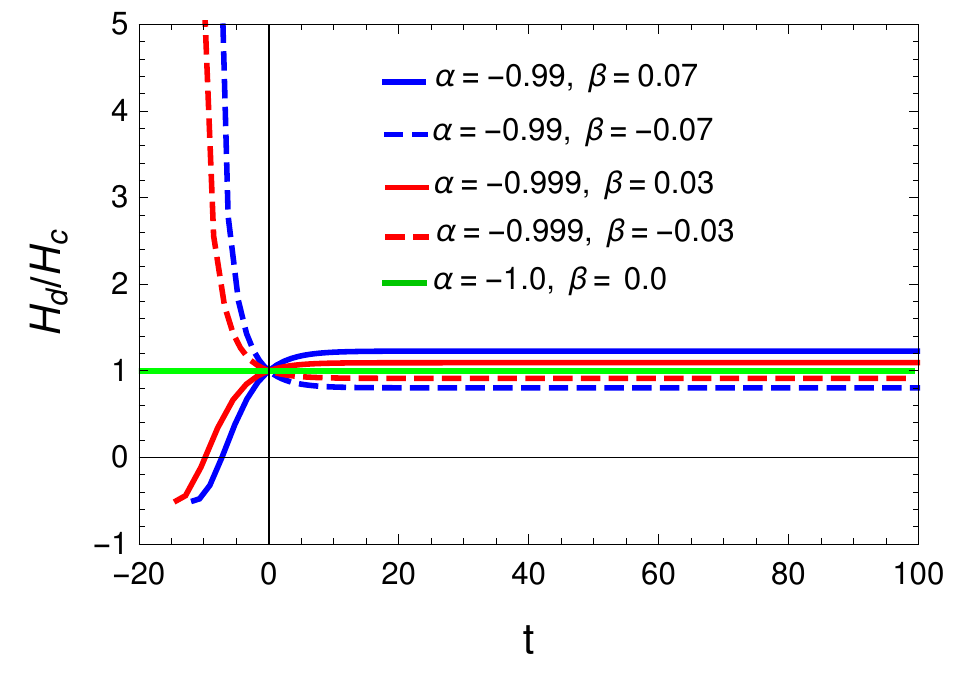}}
\caption{Plots of $H_d/H_c$ for several values of $\alpha$ and $\beta$ near $\alpha=-1$ and  $\beta=0$.
The anisotropy is large at early times, and small at late times.}
\label{fig5}
\end{figure}

Let us discuss some limits of the parameters.
For the late-time solutions in G1 and G3,
let us expand $H_c$ and $H_d$ about two equation-of-sate parameters 
at $\alpha=-1$ (to explain current observational acceleration) 
and $\beta=0$ (to have small anisotropy),
\begin{align}\label{Hcapprox}
H_c &\approx \frac{2}{3(1+\alpha)t} \left[ 1-\beta\left( 1+\frac{2\beta}{1+\alpha} \right) + \cdots \right],\\
H_d &\approx \frac{2}{3(1+\alpha)t} \left[ 1+2\beta\left( 1-\frac{\beta}{1+\alpha} \right) + \cdots \right].
\end{align}
In these parameter expansions, the anisotropy, i.e., the difference in $H_c$ and $H_d$ 
starts to appear in the second-leading order. 
The expansions of spacetime with these parameter ranges ($\alpha \approx -1$ and $\beta \approx 0$) 
may give some hints for explaining the current acceleration of the Universe with a negligible anisotropy.
However, these ranges mimic the cosmological constant, so the evolution of the Universe will not be explained 
solely by the fluid; we need other matter contents
to explain the currently observed universe.
On the contrary, at early times, the anisotropy is large in these parameter ranges (see Fig.~\ref{fig5}).
It will be interesting to investigate its cosmological effect more in detail,   
but it is beyond the scope of this work.

%%%%%%%%%%%%%%%%%%%%%%%%%%%%%%%%%%%%%%%%%%%%%%%%%%%%%%%%%%%%%%%%%%%%%%%%%%
%%%%%%%%%%%%%%%%%%%%%%%%%%%%%%%%%%%%%%%%%%%%%%%%%%%%%%%%%%%%%%%%%%%%%%%%%%

%%%%%%%%%%%%%%%%%%%%%%%%%%%%%%%%%%%%%%%%%%%%%%%%%%%%%%%%%%%%%%%%%%%%%%%%%%
%%%%%%%%%%%%%%%%%%%%%%%%%%%%%%%%%%%%%%%%%%%%%%%%%%%%%%%%%%%%%%%%%%%%%%%%%%

\section{Conclusions}
\label{sec:conclusion}

We investigated the effect of the off-diagonal stress 
tensor in the evolution of the spacetime.
(With a suitable coordinate transformation, the stress tensor can be diagonalized and the set-up can be put into the Bianchi type VII anisotropic spacetime.)
The energy-momentum tensor of the matter fluid consists of 
the energy density, the pressure, and the shear viscosity. We considered only the off-diagonal components of the shear viscosity. We considered each component of the pressure and the stress having the same value respectively
with introducing the off-diagonal spatial metric components of the same value.

It is quite natural to consider the shear viscosity in the realistic astrophysical situations such as a molecular cloud described by fluid. In cosmology, it is also natural to consider it, but the current universe exhibits very small anisotropy, so people often ignore the shear viscosity simply for convenience. In cosmological perturbation theories, it is inevitable to consider the anisotropic tensor field (representing shear viscosity) in theories higher than linear order \citep{Noh:2004bc,Hwang:2007ni, Hwang:2017oxa}, which we described in the Introduction as an example.

We assumed the simplest equation of states for which the pressure and the off-diagonal stress
are proportional to the energy density, $p =\alpha\rho$ and $\sigma =\beta\rho$. We solved the Einstein's equation with the off-diagonal spatial components of the metric turned on.
One of the metric functions is given by an implicit integral form
which we could solve numerically.
We solved the field equations for the range of parameters, $-1 \leq \alpha, \beta \leq 1$.
Depending on the values of the parameters, 
the integration of the field equation was classified into five types.

The solutions of all the classes exhibit the initially vanishing volume 
except for the pure de Sitter ($\alpha=-1, \beta=0$),
which means that the three-volume density  ${\cal V}_3$ is zero at the initial moment $t_s$.
${\cal V}_3$ increases afterwards.

The evolution pattern of the energy density $\rho$ at the early stage varies depending on the parameters.
For some cases, $\rho$ drops from infinity as the usual Friedmman universe ($\beta =0$).
For others, however, $\rho$ starts from a finite value and increases/decreases at the early stage.
At late times, $\rho$ decreases for all classes except for the pure de Sitter.
At late times, we could derive the asymptotic solutions analytically.
They exhibit that $\rho$ decreases faster than, or
at most equal to the usual Friedmann universe considering the power of $t$-dependence.

Although the three-volume density vanishes at the initial moment, ${\cal V}_3(t_s)=0$,
the curvature is not always divergent.
We analysed the energy density and the Kretshmann curvature (${\cal K}$) near the initial moment $t_s$.
We found that $\rho(t_s)=0$ and ${\cal K}(t_s)=0$ for the parameters in the range, $1+\alpha+2\beta < 0$,
so the big-bang singularity can be removed.
We observed that the weak-energy condition is violated in this range.
Note that this is achieved when the off-diagonal stress is quite significant,
i.e., the value of $\beta$ is not very close to zero.

As a whole, introducing the off-diagonal stress in fluid, 
the energy density transfers to the stress.
This makes the energy density drops faster than the usual stress-free universe (FRW),
and the expansion of the Universe becomes slower as a result.

Our model exhibits an anisotropic evolution of spacetime.
At late times, the anisotropy settles down to a constant value except for $1-\alpha-2\beta<0$.
The anisotropy is not small unless the off-diagonal stress is negligible, $\beta \approx 0$.
With $\alpha \approx -1$ and $\beta \approx 0$,
our model may explain
the acceleration of the current universe with small anisotropy,
but the effect of the anisotropy at early times needs more investigation,
with other matter components for completion.

\section*{Acknowledgements}
Authors are grateful to Hyeong-Chan Kim, Wonwoo Lee, Byungchan Kim, and Jai-chan Hwang for useful discussions.
This work was supported by the grant from the National Research Foundation
funded by the Korean government, No. NRF-2020R1A2C1013266.

\end{document}